\documentclass[numberedappendix,apj]{openjournal}

\usepackage{amsmath,amssymb,bm,mathrsfs}
\usepackage[scr=rsfs,cal=boondox]{mathalfa}
\usepackage{graphicx}

\begin{document}

\def\cals{\mathcal{S}}
\def\cali{\mathcal{I}}
\def\calq{\mathcal{Q}}
\def\calu{\mathcal{U}}
\def\calv{\mathcal{V}}

\def\d{{\rm d}}

\title{Error Propagation from Integral Moments and Spectral Derivatives of Stokes Profiles to Magnetic Observables}
\author{R. Casini}
\affiliation{NSF NCAR, HAO, Boulder, CO 80307-3000, USA}

\lefthead{Signal Error Propagation to Magnetic Observables}
\righthead{R. Casini}

\begin{abstract}
We aim to derive convenient analytic estimations of magnetic inference errors, which can be used to inform and facilitate the creation of the science-traceability matrix of a polarimetric instrument. In order to do this,
we study the general problem of the propagation of Poissonian noise in polarized light signals to derived data products of the Stokes profiles, namely \emph{integral moments} of arbitrary orders and \emph{spectral derivatives} of the intensity. In particular, we derive the errors on the longitudinal and transverse projections of the magnetic field inferred from the weak-field approximation (WFA) of the Zeeman effect.
The propagated errors on magnetic inference are traced back to the signal-to-noise ratio (SNR) of the observations. The results can be used to quantify the contributions of different environmental and instrumental conditions (spectral line width, spectral resolution, line spread function, stray light) to the overall error on the inferred quantities. Our tests indicate that the integral-moment method of inference is comparable to the local derivative in a predictable way.
\end{abstract}
\maketitle

\section{Introduction}

The error analysis for physical quantities that are inferred via
spectro-polarimetric methods, e.g., through the inversion of the
polarization signatures of atomic transitions that occur in magnetized
solar plasmas, has traditionally been conducted using numerical tests 
where inversions of simulated Stokes data are run for a large variety 
of possible physical conditions, and the success rate and range of
applicability of the inversion method are assessed through a statistical
analysis of the results \citep{Ce18}. 
To our knowledge, only a few authors have tried to put these statistical 
analyses on a solid mathematical basis \citep{AR11,MG12}, but otherwise the more
fundamental problem of tracing the error propagation from the observed
signals to the inferred physical quantities has scarcely been given due attention \cite[most notably by][]{DM12}.

On the other hand, the availability of quick estimates of magnetic-inference 
errors, and how they are conceptually connected with instrumental properties, 
such as system's throughput and spectral resolution, can be very helpful 
during the science-to-instrument requirement flowdown of spectro-polarimeters, 
whether these are being proposed for deployment to ground-based facilities or 
in space. Commonly, the science requirements of a project that aims at 
diagnosing magnetic fields in astrophysical plasmas will include a magnetic 
sensitivity target expressed either as an absolute minimum detectable field 
strength (or flux) or a relative field strength error. Having the ability 
to quickly estimate the performance of conceptual designs of 
spectro-polarimeters without the need to invest a significant amount of time 
and resources for a detailed Stokes data simulation and magnetic inversion 
exercise would greatly streamline the creation of meaningful 
science-traceability matrices.

In this paper we want to investigate this problem by focusing on 
two principal derived products of the Stokes profiles
often used for magnetic inference, which are the integral moments of
Stokes profiles, which are preferred observables for coarsely sampled 
line profiles, especially of very dim targets (e.g., the forbidden 
emission lines formed in the solar corona; \citealt{To08}), and the
\emph{weak-field approximation} \cite[WFA;][]{LL04,Ce18}, which relates the Stokes 
polarization signals to the spectral derivative of the intensity, and 
which is often used as a quick-look diagnostics of the Zeeman effect, 
when sufficiently resolved Stokes profiles are available.

In Sects.~\ref{sec:moments} and \ref{sec:deriv} we analyze these two main data products, and derive algebraic estimators for a fast yet robust quantification of the
inference errors. In Sect.~\ref{sec:comparison} we draw some parallels
and highlight differences between the two types of data products and 
approaches, and finally we provide some basic conclusions about our 
findings.

Throughout this work we make repeated use of linearized error propagation in the presence of both uncorrelated and correlated quantities. In order to facilitate the reader in following the development of Sects.~\ref{sec:moments} and \ref{sec:deriv}, we present without proof some relevant and recurring formulas for error propagation in the next short section.

\section{Linearized Error Propagation}
\label{sec:errprop}

Linearized error propagation is the tool commonly adopted to derive estimations of noise errors on (inferred) quantities $F,G,\ldots$ that are general functions of a set of $n$ (observed) random variables $\{x_i\}_{i=1}^n$, based on some assumed physical model. The variables $x_i$ may or may not be independent (i.e., uncorrelated) of each other.

The formulation is based on the general definition of ``linearized'' \emph{covariance} between two quantities $F,G$ (see, e.g., \citealt{KS77v1}, Ch.~10)
\begin{equation} \label{eq:EPgen}
    \mathrm{Cov}(F,G)=\sum_{i,j=1}^n
    \frac{\partial F}{\partial x_i}
    \frac{\partial G}{\partial x_j}\,
    \mathrm{Cov}(x_i,x_j)\;.
\end{equation}
The diagonal case $F=G$ corresponds to the usual definition of \emph{variance}
\begin{equation} \label{eq:EPvar}
\sigma^2(F)\equiv\mathrm{Cov}(F,F)
=\sum_{i,j=1}^n
    \frac{\partial F}{\partial x_i}
    \frac{\partial F}{\partial x_j}\,
    \mathrm{Cov}(x_i,x_j)\;,
\end{equation}
and in the particular case when \emph{all} random variables are uncorrelated, so that $\mathrm{Cov}(x_i,x_j)=\delta_{ij}\,\sigma^2(x_i)$, the well-known \emph{diagonal} error-propagation formula ensues,
\begin{equation} \label{eq:EPvardiag}
\sigma^2(F)=\sum_{i=1}^n
    \left(\frac{\partial F}{\partial x_i}\right)^{\!2}
    \sigma^2(x_i)\;.
\end{equation}

The above relations can be cast in the form of \emph{relative} errors, by noting that
\begin{eqnarray} \label{eq:EPrelgen}
\frac{\mathrm{Cov}(F,G)}{FG}
&=&\sum_{i,j=1}^n
    \left(\frac{x_i}{F}\frac{\partial F}{\partial x_i}\right)\!
    \left(\frac{x_j}{G}\frac{\partial G}{\partial x_j}\right)
    \frac{\mathrm{Cov}(x_i,x_j)}{x_i x_j} \nonumber \\
&=&\sum_{i,j=1}^n
    \frac{\partial \ln F}{\partial \ln x_i}
    \frac{\partial \ln G}{\partial \ln x_j}\,
    \frac{\mathrm{Cov}(x_i,x_j)}{x_i x_j}\;.
\end{eqnarray}
In particular, for the most common case of uncorrelated variables, Eq.~(\ref{eq:EPvardiag}) becomes
\begin{equation} \label{eq:EPreldiag}
\frac{\sigma^2(F)}{F^2}=\sum_{i=1}^n
    \left(\frac{\partial\ln F}{\partial\ln x_i}\right)^{\!2}
    \frac{\sigma^2(x_i)}{x_i^2}\;.
\end{equation}

The above formulas summarize the method of linearized error propagation, and they will be applied extensively in the derivation of the main results presented in the following Sects.~\ref{sec:moments}
and \ref{sec:deriv}. However, it is important to remark that the application of such formulas in the case of estimators $F$ that depend non-linearly on the random variables $x_i$ requires in practice that the relative variance $\sigma^2(x_i)/x^2_i$ be small, so that $F$ is
approximately linear over the range of the fluctuations of $x_i$. This restriction does not apply to quantities that enter the expression of the estimator linearly, which then may carry arbitrarily large errors.

\section{Integral moments of Stokes signals}
\label{sec:moments}

Let $S(\lambda)$ be a continuous intensity signal affected by a stationary random noise
$\sigma_S$. For photon-noise limited observations, when the \emph{measured} signals $S_\lambda$ are
expressed in photon counts, we can assume $\sigma_S(\lambda)\approx\sqrt{S_\lambda}$. Operationally speaking, $S_\lambda\equiv\langle S(\lambda)\rangle$, where the average operation represents the measuring process itself, which implies an integration of the signal $S(\lambda)$ over the elemental spectral, spatial, and temporal intervals of the measuring apparatus.

The spectral integral moment of order $m$ of the signal $S(\lambda)$ over a bandwidth $(\lambda_a,\lambda_b)$, expressed as the photon count per unit wavelength collected during the signal acquisition, is defined as\footnote{Throughout this work we use capital script letters to indicate moments of the Stokes signals. Hence, the 0th-order of Stokes $I(\lambda)$ will be indicated with $\cali_0$, the 1st-order moment of $V(\lambda)$ with $\calv_1$, etc.}
\begin{equation} \label{eq:integral}
\cals_m\equiv\int_{\lambda_a}^{\lambda_b}\d\lambda\;(\lambda-\lambda_0)^m\,S(\lambda)\;,
\end{equation}
where $\lambda_0$ is the reference wavelength for the moments (for a
line profile, typically the line center).
We want to determine the propagated noise from the signal to its
moments, which in general is expected to depend on the spectral 
resolution of the profile, as we are going to show below.

If $\delta\lambda$ is the spectral sampling interval, which we can 
assume to be homogeneous across the spectral range, with the above definitions the measured signal is already integrated over such an interval,
and so the transformation from the continuous expression 
(\ref{eq:integral}) to the corresponding discrete case of 
a sampled signal is simply
\begin{eqnarray} \label{eq:sampling}
\cals_m\approx\sum_{i=1}^N (\lambda_i-\lambda_0)^m\,S_i\;,\qquad
N&=&\frac{\lambda_b-\lambda_a}{\delta\lambda}\;,\quad
\lambda_i\equiv\lambda_a+\bigl(i-\textstyle\frac{1}{2}\bigr)\delta\lambda\;,
\end{eqnarray}
where $S_i$ is the integrated signal $S(\lambda_i)$ over the pixel width $\delta\lambda$.
If now $R$ is the spectral resolving power of the observation, \emph{assuming 
critical sampling of the resolution spectral element}, we additionally can write 
$\delta\lambda\approx\lambda_0/2R$.

Because the acquisition of the $S_i$ signals implies a set of independent measurements, for photon-noise limited observations we can write (see Eq.~(\ref{eq:EPvardiag}))
\begin{equation} \label{eq:noise}
\sigma^2(\cals_m)\approx\sum_{i=1}^N
	(\lambda_i-\lambda_0)^{2m}\,\sigma^2(S_i)
\approx\sum_{i=1}^N 
	(\lambda_i-\lambda_0)^{2m}\,S_i\;,
\end{equation}
leading to the fundamental result
\begin{equation}  \label{eq:noise1}
\sigma^2(\cals_m)\approx\cals_{2m}\;,
\end{equation}
in virtue of the definition (\ref{eq:sampling}) of spectral moments. It is important to remark that the adopted approximation $\sigma^2(S_i)\approx S_i$, which holds for a Poisson distributed signal measured at the detector, is here being implicitly extended to the Stokes intensity of a generally polarized signal. On the other hand, the inferred Stokes intensity is always the result of a polarimetric analysis involving weighted combinations of the actual measured signals from the target, and such an analysis generally breaks the Poissonian character of the inferred intensity. In App.~\ref{app:covariances} we clarify the assumptions under which Eq.~(\ref{eq:noise1}), and all following relations relying on it, can be extended to the inferred Stokes intensity of a target.

In order to apply the error propagation formulas of Sect.~\ref{sec:errprop} to the problem of estimating  magnetic inference errors in astrophysical applications, we need to provide a formation model for the polarization signals of a spectral line in a magnetized plasma. For many typical problems of interest in astrophysics, a treatment of the Zeeman-effect signatures in the Stokes parameters of the line is sufficiently representative, and since magnetic inference errors are more critical at low field strengths, here we target specifically the regime of applicability of the WFA of the Zeeman effect.

We start recalling the expression of the WFA of Stokes $V$ \citep{LL04} for the determination of the LOS component of the magnetic field, which allows us to write
\begin{equation} \label{eq:WFA_V}
V(\lambda)\approx-\beta\lambda_0^2\,\frac{\d}{\d\lambda}I(\lambda)\;,
\end{equation}
where $\lambda_0$ is the line central wavelength, and $\beta$ is a 
quantity with the dimensions of an inverse length, which depends on the 
magnetic field direction and strength, and on the sensitivity of the line 
to the Zeeman effect (via the so-called effective Land\'e factor
$\bar{g}$). The explicit expression of $\beta$ is given in terms of the 
Larmor frequency $\nu_B$ of the applied field:
\begin{equation} \label{eq:beta}
\beta\equiv\bar{g}\,\frac{\nu_B}{c}\cos\Theta_B\;,
\end{equation}
where $\Theta_B$ is the angle between the magnetic field vector and the
line-of-sight (LOS).

It is important to remark that the approximation (\ref{eq:WFA_V}) is strictly valid in the limit of small fields (as well as of negligible fine structure of the transition), so the modification of the intensity profile from the energy separation of the line components can be neglected. The reason for this is that the WFA derives from a perturbation expansion of the radiative transfer equation of polarized radiation, where the small parameter of the perturbation corresponds to the ratio between the energy span of the transition (due to the external fields and the fine structure of the ion) and the thermal width of the line \cite[see, e.g.,][and references therein]{LL73,CL96,AC26}. Thus, rigorously speaking, the intensity profile appearing in the RHS of Eq.~(\ref{eq:WFA_V}) should be the one corresponding to the zeroth-order approximation, only accounting for the thermal broadening of the line.

For the following development, we have implemented the WFA exactly, in order to track exclusively the effects of the propagation of Poissonian noise. Thus, the intensity profile $I(\lambda)$ entering Eq.~(\ref{eq:WFA_V}), as well as Eq.~(\ref{eq:WFA_Q}) later in this section, is a pure Gaussian profile with a given thermal width and the additional smearing by a spectral PSF, which corresponds precisely to the lowest order of the WFA of the intensity profile.

Using the approximation (\ref{eq:WFA_V}) to define the 1st-order 
moment of $V(\lambda)$ according to Eq.~(\ref{eq:integral}), we obtain
\begin{eqnarray} \label{eq:wfa-moment_V}
\calv_1
&\approx&-\beta\lambda_0^2\int_{\lambda_a}^{\lambda_b}
	\d\lambda\;(\lambda-\lambda_0)\,\frac{\d}{\d\lambda}I(\lambda) 
=-\beta\lambda_0^2\left[
	(\lambda-\lambda_0)\,I(\lambda)\Bigr|_{\lambda_a}^{\lambda_b}
	-\int_{\lambda_a}^{\lambda_b}\d\lambda\;I(\lambda)\right]
\nonumber \\
&\approx&\beta\lambda_0^2\left[\cali_0
	-(\lambda_b-\lambda_a)\,I^\mathrm{b}\right]\;,
\end{eqnarray}
where in the last approximation we 
assumed that the wavelengths $\lambda_{a,b}$ 
lie far enough from the line center that the signal is completely
dominated by a continuous background, assumed to be \emph{constant and
non-polarized} across the spectral domain, i.e.,
$I(\lambda_a)\sim I(\lambda_b)\approx I^\mathrm{b}$. We note that the 
relation (\ref{eq:wfa-moment_V}) applies equally to emission and absorption lines, as the contribution of any continuum is automatically subtracted. However, both the continuum level and the width of the spectral window $(\lambda_a,\lambda_b)$ over which the moment is calculated bear on the error on the magnetic inference, as shown below.

Equation~(\ref{eq:wfa-moment_V}) expresses the well-known result that the 
longitudinal field is estimated by the 1st-order moment of Stokes $V$, 
according to
\begin{equation} \label{eq:magn.diagn}
\beta\approx\frac{1}{\lambda_0^2}\,\frac{\calv_1}{\cali_0-\cali_0^\mathrm{b}}\;,
\end{equation}
where, according to Eq.~(\ref{eq:sampling}), $\cali_0^\mathrm{b}\approx
N\delta\lambda\, I^\mathrm{b}$.
Error propagation must take into account the fact that the observables $\calv_1$ and $\cali_0$ are not statistically independent, as they are built from the same set of signals. For example, in the case of an ideal, dedicated $(I,V)$ dual-beam polarimeter, if we indicate with $v_i^\pm$ the signals of the two beams at the sampled wavelength $\lambda_i$, both $I_i$ and $V_i$ are derived from those signals, being $I_i=v_i^++v_i^-$ and $V_i=v_i^+-v_i^-$. In contrast, according to Eq.~(\ref{eq:wfa-moment_V}), the background level $I^\mathrm{b}$, and hence $\cali_0^\mathrm{b}$, is determined exclusively by the values of the intensity at the edges of the bandwidth $(\lambda_a,\lambda_b)$, which carry no weight in the integrals defining $\calv_1$ and $\cali_0$. Thus, when we calculate these moments through Eq.~(\ref{eq:sampling}), it is legitimate to
assume that the background term has been evaluated from values outside of that bandwidth (i.e., from the neighboring continuum), so that any correlation between $\cali_0^{\rm b}$ and $\calv_1$
or $\cali_0$ can be neglected.
Using Eq.~(\ref{eq:EPvar}) applied to Eq.~(\ref{eq:magn.diagn}) as a function of the quantities $\calv_1$, $\cali_0$, and $\cali_0^\mathrm{b}$ then gives
%
\begin{eqnarray} \label{eq:sigmabet.tmp}
\sigma^2(\beta)
&\approx&\frac{1}{\lambda_0^4}\frac{\sigma^2(\calv_1)}{(\cali_0-\cali_0^\mathrm{b})^2}
	+\frac{1}{\lambda_0^4}\,\calv_1^2\,
	\frac{\sigma^2(\cali_0)+\sigma^2(\cali_0^\mathrm{b})}{(\cali_0-\cali_0^\mathrm{b})^4}
    -\frac{1}{\lambda_0^4}\,\calv_1\,
	\frac{2\,\mathrm{Cov(\calv_1,\cali_0)}}{(\cali_0-\cali_0^\mathrm{b})^3}
\nonumber \\
&\approx&\frac{1}{\lambda_0^4}\frac{\sigma^2(\calv_1)}{(\cali_0-\cali_0^\mathrm{b})^2}
	+\beta^2\,\frac{\cali_0+\cali_0^\mathrm{b}}{(\cali_0-\cali_0^\mathrm{b})^2}
    -\frac{1}{\lambda_0^2}\,\beta\,\frac{2\,\mathrm{Cov(\calv_1,\cali_0)}}{(\cali_0-\cali_0^\mathrm{b})^2}\;,
\end{eqnarray}
where in the second row we used again Eq.~(\ref{eq:magn.diagn}), as well as Eq.~(\ref{eq:noise1}) for $m=0$.

\begin{figure}[t!]
\centering
\includegraphics[width=.495\linewidth]{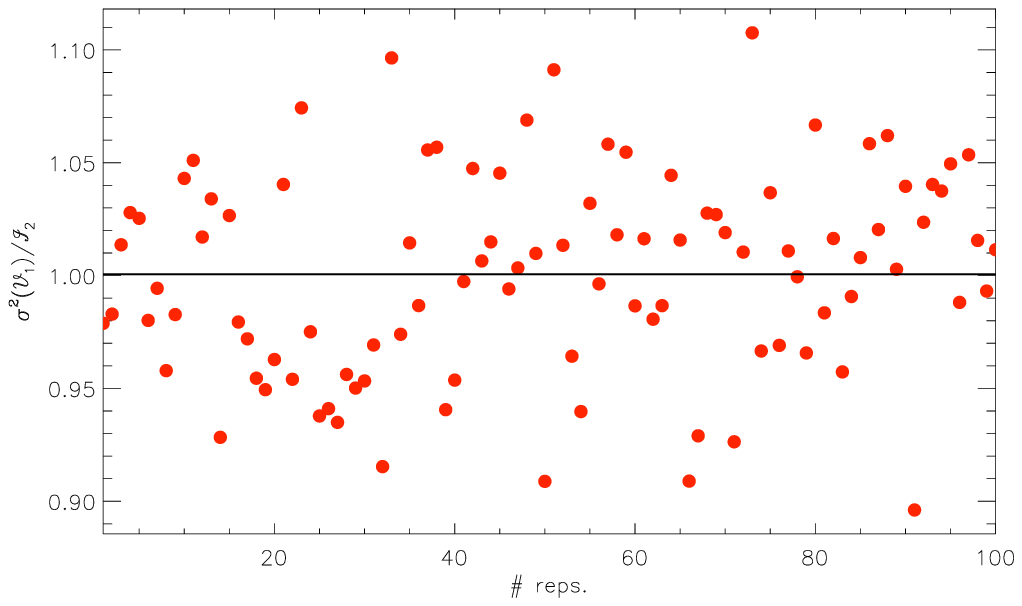}
\includegraphics[width=.495\linewidth]{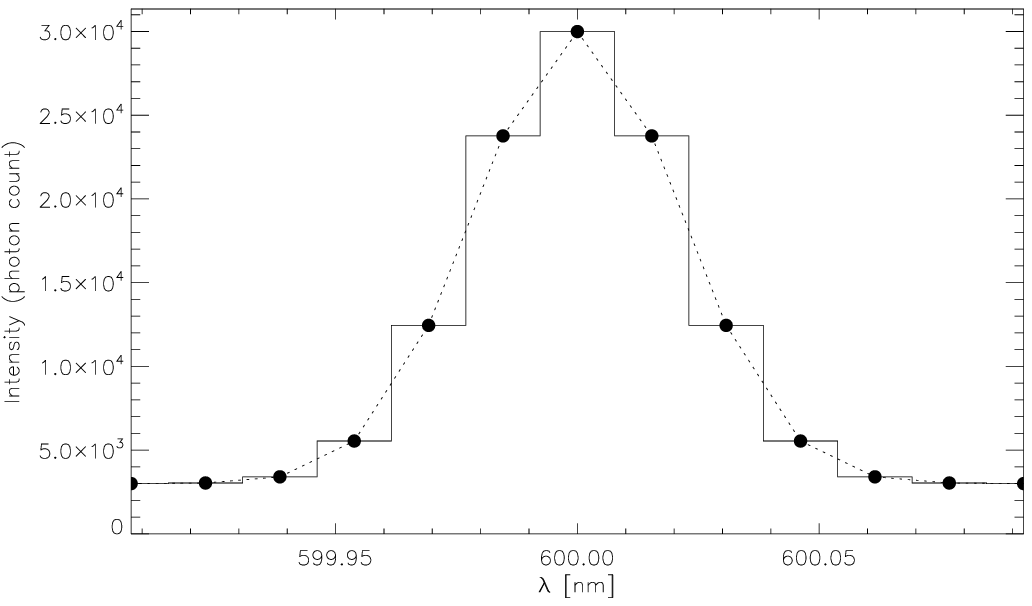}
\caption{\small \label{fig:v1-s2}
\emph{Left:} numerical verification of the validity ``in the mean'' of the moment 
relation (\ref{eq:v2mom}) in the presence of Poissonian noise. 
Each red point corresponds to the variance over 1000 random realizations 
of the noise. The horizontal black line marks the mean of the distribution
of this variance over 100 repetitions of the test. For the non-noised 
Stokes profiles, we considered an emission line with a Gaussian shape, with a peak amplitude of 30000 counts, including a 10\% background continuum relative to the peak, and a FWHM of 0.05\,nm (right),
and a weak magnetic field of 100\,G to compute Stokes $V$. The 
spectral sampling adopted for this test (13 points across the full 
spectral range) corresponds to a resolving power $R\approx 20000$, and the FWHM of the profile includes the instrument spectral PSF. A fictitious integration time was assumed in order to accumulate approximately 30000 photon counts, as shown in the right panel.}
\end{figure}

In order to evaluate $\sigma^2(\calv_1)$, we must specify the
demodulation scheme relating the Stokes $V$ observable to the
modulated intensity signals acquired by the instrument \cite[e.g.,][]{Ca26}. 
%
%
From the definition of the polarization \emph{modulation efficiencies} $(\epsilon_I,\epsilon_Q,\epsilon_U,\epsilon_V)$ of a full-Stokes polarimeter \citep{dTIC00}, we can immediately write (see also App.~\ref{app:covariances})
\begin{equation} \label{eq:v2mom}
\sigma^2(\calv_1)\approx
\frac{\epsilon_I^2}{\epsilon_V^2}\,\sigma^2(\cali_1)
\approx r_V^2\,\cali_2\;,
\end{equation}
where we recalled Eq.~(\ref{eq:noise1}), and also defined $r_S=\epsilon_I/\epsilon_S$ for $S=Q,U,V$ (see Eq.~(\ref{eq:rSdef})).
The validity of this approximation can directly be tested numerically.
Figure~\ref{fig:v1-s2} (left) shows the statistical distribution of the 
ratio between the two sides of Eq.~(\ref{eq:v2mom}), assuming an optimally efficient $(I,V)$ polarimeter, so that $\epsilon_V=\epsilon_I$. We assumed a Gaussian profile
shape of the intensity profile using 13 wavelength sample points across the spectral range (right),
for an effective resolving power of 20000, and a longitudinal 
magnetic field of 100\,G for the calculation of Stokes $V$ (see caption
for more details). This figure demonstrates that the expectation value of 
the ratio $\sigma^2(\calv_1)/\cali_2$ tends to 1 as expected.

The type of modulation scheme also determines the form of $\mathrm{Cov}(\calv_1,\cali_0)$. For an optimally efficient $(I,V)$ dual-beam polarimeter, we easily find $\mathrm{Cov}(\calv_1,\cali_0)\approx\calv_1$.\footnote{This result is formally derived in App.~\ref{app:covariances}, but it can more directly be arrived at by noting that $\calv_1=\sum_i(\lambda_i-\lambda_0)\,(v^+_i-v^-_i)$ and $\cali_0=\sum_i(v^+_i+v^-_i)$, where $v^\pm_i$ are the signals of the two beams, which here act as the independent variables in Eq.~(\ref{eq:EPvar}).} In fact, such a result is more general, and in particular it holds for any optimal modulation scheme (see App.~\ref{app:covariances}). Therefore, we are going to adopt such a result in Eq.~(\ref{eq:sigmabet.tmp}), which after some simple algebra becomes
\begin{equation} \label{eq:sigmabeta}
\sigma^2(\beta)\approx
\frac{r_V^2}{\lambda_0^2(\cali_0-\cali_0^\mathrm{b})}
	\left(\frac{1}{\lambda_0^2}
\frac{\cali_2^\mathrm{l}+\cali_2^\mathrm{b}}{\cali_0-\cali_0^\mathrm{b}}
-\frac{\beta^2\lambda_0^2}{r_V^2}\,\frac{\cali_0-3\cali_0^\mathrm{b}}{\cali_0-\cali_0^\mathrm{b}}\right)\;,
\end{equation}
where we also used Eq.~(\ref{eq:v2mom}), and we explicitly separated the contributions of the line (l) and the
background in $\cali_2$.

\begin{figure}[t!]
\centering
\includegraphics[width=.495\linewidth]{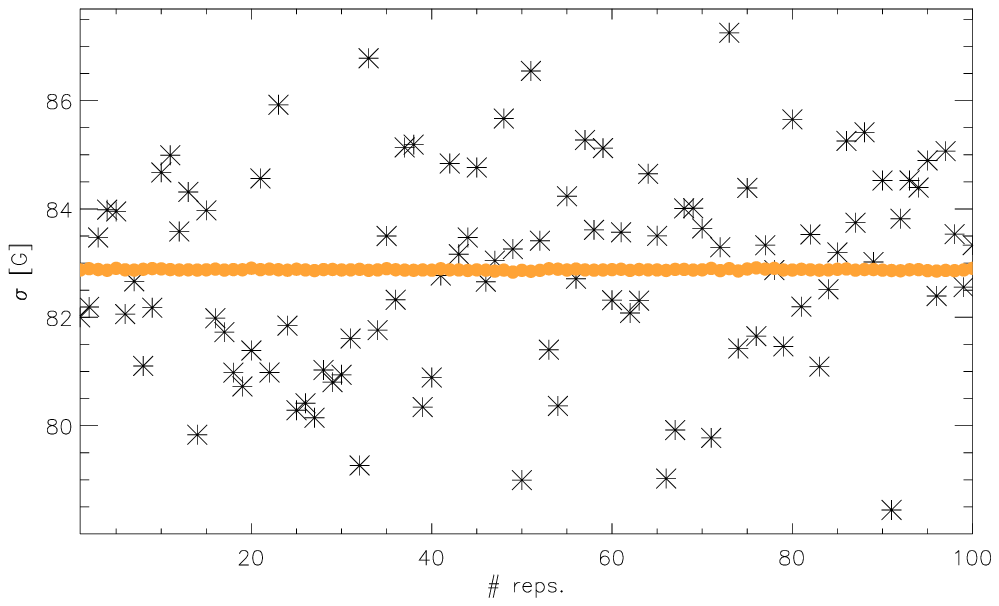}
\includegraphics[width=.495\linewidth]{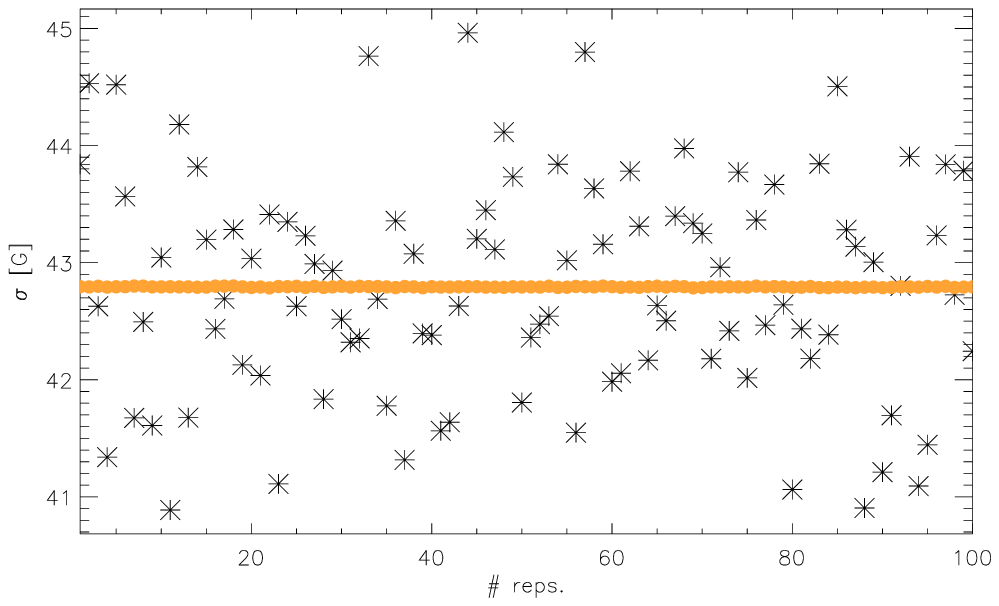}
\caption{\small \label{fig:sgmBmoment}
Numerical verification of the validity of the estimator 
(\ref{eq:error-beta}) of the magnetic inference variance, using the 
same Stokes $I$ and $V$ profiles and testing conditions of 
Fig.~\ref{fig:v1-s2} (left), and for a resolving power of 100000
(right), assuming that the integration time has been increased so as to keep the photon count per spectral bin the same as in the lower resolution case. 
Each star symbol represents the standard deviation of the magnetic 
inference using Eq.~(\ref{eq:magn.diagn}) over 1000 random realizations 
of the noise. The orange dots represent the standard deviation estimator 
from Eq.~(\ref{eq:error-beta}). 
Both error distributions yield approximately the same means (82.9\,G for 
the lower resolution case and 42.8\,G in the higher resolution case) over 
100 repetitions of the test.}
\end{figure}

For a Gaussian profile with ($e$-folding) thermal Doppler halfwidth
\begin{equation}
\Delta_T=\lambda_0\frac{v_T}{c}\equiv\lambda_0\sqrt{\frac{2kT}{mc^2}}\;,
\end{equation}
we have explicitly
\begin{displaymath}
\frac{\cali_2^\mathrm{l}}{\cali_0-\cali_0^\mathrm{b}}
	=\frac{\Delta_T^2}{2}=\frac{1}{2}\,\lambda_0^2\,\frac{v_T^2}{c^2}\;.
\end{displaymath}
In the additional presence of instrumental broadening by a spectral PSF
that can be approximated by a Gaussian profile with $e$-folding halfwidth
$\Delta_{\rm PSF}$, we can further write
\begin{equation} \label{eq:gaussian2moment}
\frac{\cali_2^\mathrm{l}}{\cali_0-\cali_0^\mathrm{b}}
	=\frac{\Delta_T^2+\Delta_{\rm PSF}^2}{2}
=\frac{1}{2}\,\lambda_0^2\,\frac{v_T^2}{c^2}+\frac{1}{2}\,\Delta_{\rm PSF}^2\;.
\end{equation}
%
%
%
Next we observe that we can write
\begin{displaymath}
\cali_2^\mathrm{b}
\approx \delta\lambda\,I^\mathrm{b}
	\sum_{i=1}^N(\lambda_i-\lambda_0)^2\
\approx \cali_0^\mathrm{b}\,\frac{1}{N} \sum_{i=1}^N(\lambda_i-\lambda_0)^2\;.
\end{displaymath}
Thus, Eq.~(\ref{eq:sigmabeta}) becomes
\begin{eqnarray} \label{eq:error-beta}
\sigma^2(\beta)\approx
\frac{r_V^2}{2\lambda_0^2(\cali_0-\cali_0^\mathrm{b})}\biggl(\frac{v_T^2}{c^2}
&+&\frac{\Delta_\mathrm{PSF}^2}{\lambda_0^2}
+\frac{\overline{(\lambda-\lambda_0)^2}}{\lambda_0^2}
	\frac{2\cali_0^\mathrm{b}}{\cali_0-\cali_0^\mathrm{b}} 
-2\,\frac{\beta^2\lambda_0^2}{r_V^2}\,
	\frac{\cali_0-3\cali_0^\mathrm{b}}{\cali_0-\cali_0^\mathrm{b}}\biggr)\;.
\end{eqnarray}
In particular, if we recall Eq.~(\ref{eq:sampling}), and assume for simplicity $\lambda_0=(\lambda_a+\lambda_b)/2$, we find explicitly
\begin{equation} \label{eq:avglambda}
\overline{(\lambda-\lambda_0)^2}
\equiv \frac{1}{N} \sum_{i=1}^N(\lambda_i-\lambda_0)^2
=\frac{N^2-1}{12}\,\delta\lambda^2
\approx\frac{(\lambda_b-\lambda_a)^2}{12}\;,
\end{equation}
where the last approximation improves rapidly with $N$.

\begin{figure}[!t]
\centering
\includegraphics[width=.495\linewidth]{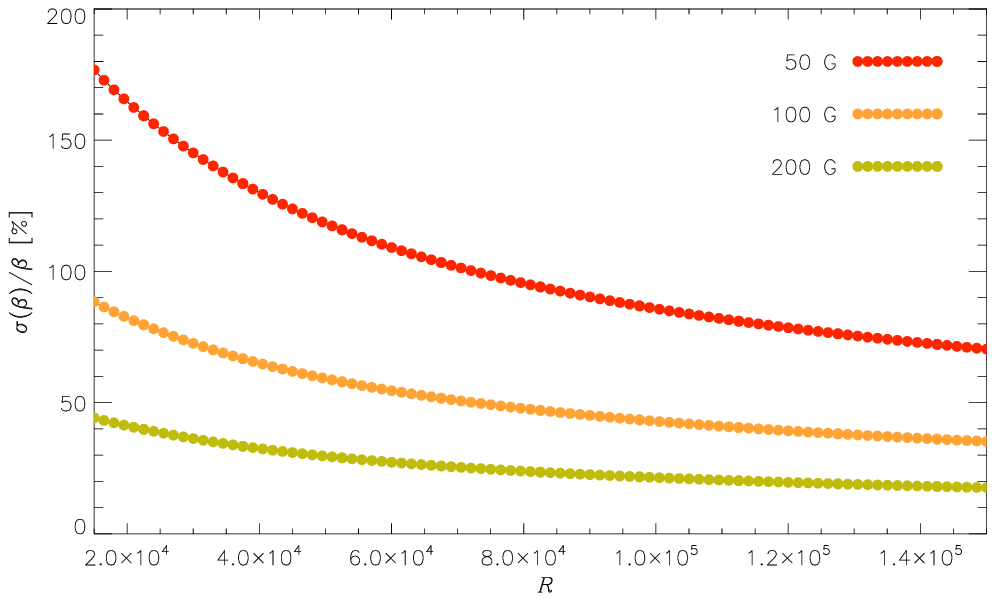}
\includegraphics[width=.495\linewidth]{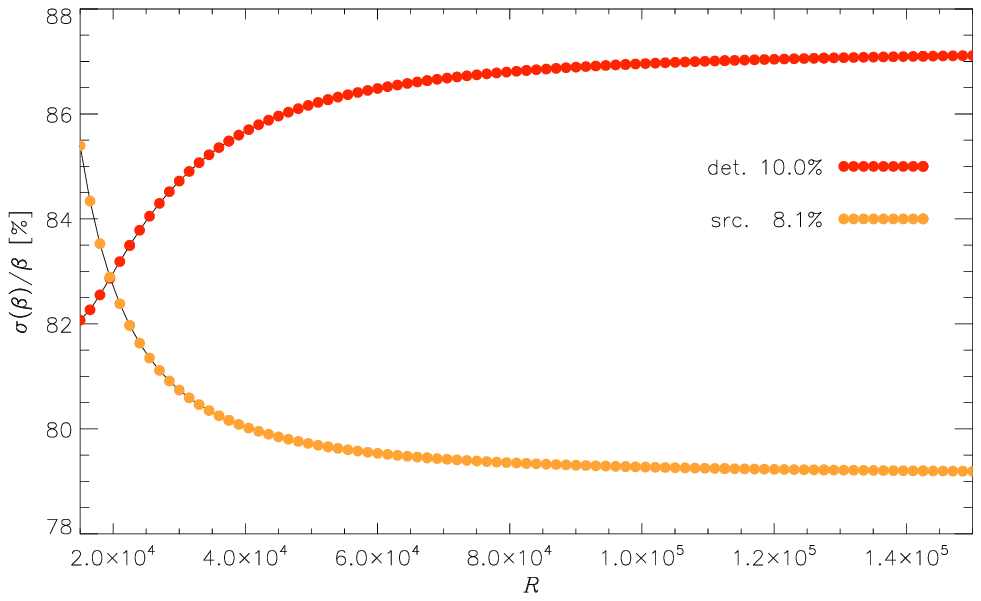}
\caption{\label{fig:test_resol}
The relative error $\sigma(\beta)/\beta$ derived from
Eq.~(\ref{eq:error-beta}) as a function of $R$. \emph{Left:} assuming a \emph{fixed
photon count} per spectral bin, as in the case of Fig.~\ref{fig:sgmBmoment}, for three different values of the magnetic field. 
On the $B=100$\,G curve, it is possible to match the noise values corresponding to the two cases shown in Fig.~\ref{fig:sgmBmoment} at $R\approx20000$ and $R\approx100000$. In the range of $R$ values shown here, the absolute errors are practically the same for the three values of the magnetic strength, so the relative errors are shown in order to distinguish the three cases. \emph{Right:} assuming a \emph{fixed integration time} for $B=100$\,G, and  two different assumptions for the background/signal ratio. In the first case (red dots), for all values of 
$R$ the background is set to the same 1/10 ratio of the peak flux at line center \emph{as measured at the detector}; in the second case (orange dots) the background is still set to a constant ratio of the peak flux, but this time corresponding to the \emph{unbroadened line at the source}, so that the background level is independent of 
$R$ by construction. For this second case, the background ratio is set to 8.1\%, so as to correspond to a 10\% ratio at the detector for 
$R\approx20000$. This is the point where the two curves cross, and the realized line profile at the detector in both cases matches the one shown in the right panel of Fig.~\ref{fig:v1-s2}.}
\end{figure}

We note that all the quantities within parentheses in
Eq.~(\ref{eq:error-beta}) are dimensionless, thus $\sigma(\beta)$ has 
the expected dimensions of an inverse length, since
$\cali_0-\cali_0^\mathrm{b}$ in the denominator of the common factor is just 
the total number of photons collected over the $(\lambda_a,\lambda_b)$ 
bandwidth of the spectral line, after any background subtraction.
We also note that the ratio $\lambda_0/\Delta_{\rm PSF}$ is directly related to the resolving power $R$ of the measurements. Because the halfwidth at half maximum 
(HWHM) of a Gaussian and the $e$-folding halfwidth $\Delta$ are related via 
$\mathrm{HWHM}=\sqrt{\ln2}\,\Delta$, under the usual 
assumption that the spectral resolution corresponds to the critical sampling of the PSF by the detector, we have 
$R\approx\lambda_0/(2\sqrt{\ln2}\,\Delta_\mathrm{PSF})
\approx 0.60\,\lambda_0/\Delta_\mathrm{PSF}$.

The validity of Eq.~(\ref{eq:error-beta}) as an estimator
of the variance of the inferred magnetic strength $\beta$ can numerically 
be tested by creating many ensembles of Stokes $I$ and $V$ profiles with
random Poissonian noise, and using Eq.~(\ref{eq:magn.diagn}) to
determine the corresponding distributions of the inferred magnetic 
strengths. 
Typical results are shown in Fig.~\ref{fig:sgmBmoment}, where we used the
same thermal and magnetic model for the line profile as in 
Fig.~\ref{fig:v1-s2}. The star symbols in both plots represent the 
standard deviation of the magnetic inference from Eq.~(\ref{eq:magn.diagn}) 
over 1000 random realizations of the noise, and the plots shows the 
distribution of those errors over 100 repetitions of the same experiment. 
The orange dots represent instead the values predicted by the error 
estimator of Eq.~(\ref{eq:error-beta}). Both error distributions have 
practically the same expectation value, demonstrating the robustness 
of the error estimator (\ref{eq:error-beta}). 
Specifically, we found $B=100.6\,\rm G\pm 82.9\,G$ in the lower resolution
case (left panel), and $B=99.9\,\rm G\pm 42.8\,G$ in the higher resolution
case (right panel).
The error reduction by approximately a factor 2 of the magnetic inference error, in this limit of relatively small field strengths, is principally determined by the increased SNR of the observations as a consequence of the longer (${\sim}\,4\times$) integration time necessary  
to preserve the photon count per pixel at the higher spectral resolution (see also caption of Fig.~\ref{fig:sgmBmoment}).

This effect is more comprehensively demonstrated in Fig.~\ref{fig:test_resol}, where we present different manifestations of the dependence of the inference error (\ref{eq:error-beta}) on the spectral resolution of the profile. The left panel realizes the same observing conditions as in Fig.~\ref{fig:sgmBmoment}, and the observed decrease of the error with $R$ is mainly caused by the increasing SNR implied by the constant photon count per spectral bin assumed for these plots. The two cases of Fig.~\ref{fig:sgmBmoment} correspond in fact to the values at $R\approx20000$ and $R\approx100000$ on the curve for $B=100$\,G. The relative error $\sigma(\beta)/\beta\equiv\sigma(B)/B$ is seen to scale nearly exactly as $1/B$, implying that the absolute error $\sigma(B)$ in this field-strength regime is largely independent of the magnetic strength, because of the last addendum in Eq.~(\ref{eq:error-beta}) only bringing about a small correction compared to the other contributions.

The right panel of Fig.~\ref{fig:test_resol} shows instead the dependence of $\sigma(\beta)/\beta$ on $R$ for a field strength of 100\,G and a fixed integration time of the observations, so the total integrated fluxes $\cali_0\pm\cali_0^\mathrm{b}$ are largely independent of $R$, except perhaps in the case of coarse sampling of the line profile, when the PSF contributes significantly to the line width. In this case, the error dependence is mostly determined by the background term, and the two curves represent two different realizations of the background (see caption for details). Otherwise, the quantities $\cali_0\pm\cali_0^\mathrm{b}$ are mainly determined by the spectral line, including any background signal, and the throughput of the instrument, but are largely independent of the spectral resolution of the measurements. This is the reason why both curves remain within a few percent of their common value at $R\approx 20000$, which also corresponds to the same value on the $B=100$\,G curve in the left panel.

%
To see this more clearly, we use Eq.~(\ref{eq:error-beta}) to rewrite
\begin{equation} \label{eq:error-beta_alt}
\frac{\sigma^2(\beta)}{\beta^2}\approx
\frac{r_V^2}{\cali_0-\cali_0^\mathrm{b}}\biggl[\frac{v_T^2/c^2+f(R)}{2\beta^2\lambda_0^2}
-\frac{1}{r_V^2}\,
	\frac{\cali_0-3\cali_0^\mathrm{b}}{\cali_0-\cali_0^\mathrm{b}}\biggr]\;,
\end{equation}
where we also defined (cf.\ Eq.~(\ref{eq:avglambda}))
\begin{equation}
f(R)=\frac{1}{4\ln2}\,\frac{1}{R^2}+\frac{(\lambda_b-\lambda_a)^2}{12\,\lambda_0^2}\frac{2\cali_0^\mathrm{b}}{\cali_0-\cali_0^\mathrm{b}}\;.
\end{equation}
For $R\gg 1$, we see that $f(R)$ becomes completely dominated by the presence of a background. The asymptotic trends shown in the right panel of Fig.~\ref{fig:test_resol} are thus principally determined by the ``saturation'' of the numerator $v_T^2/c^2+f(R)$ in Eq.~(\ref{eq:error-beta_alt}), with the initial variation at relatively low $R$ occurring when the two terms in that numerator are of comparable size. Only in the absence of a background we find that $\sigma(\beta)/\beta$ becomes simply proportional to $v_{\rm T}/(\beta\lambda_0 c)$ in the limit $R\,v_{\rm T}/c\gg 1$.

Equation~(\ref{eq:error-beta_alt}) also illustrates
how the error on the magnetic inference depends on the wavelength
of the spectral diagnostics. Because the second addendum within brackets is typically of order unity, while $v_{\rm T}/(\beta\lambda_0 c)\gg 1$ in virtue of the WFA, the first addendum rapidly dominates the relative error, manifesting at once the inverse scaling of the relative error with the magnetic strength (see discussion of the left panel of Fig.~\ref{fig:test_resol}), as well as the well-known scaling of the error with $1/\lambda_0$. 


A similar analysis can be conducted for the linear polarization induced by the Zeeman effect in the weak-field limit. The WFA in this case is subject to stricter conditions for its applicability across the entire spectral range of a polarized spectral line \cite[see][\S 9.6]{LL04}, which is a necessary condition in order to establish an integral-moment based magnetic diagnostics. However, under the rather general assumption of a non-saturated spectral line, such that the line shape can still be assumed to be approximately a Voigt function (or a Gaussian, to match the particular conditions we have examined in the case of Stokes $V$), a form of WFA of the linear polarization that is applicable to the full spectral domain of the line can be proposed, which is proportional to the 2nd-order derivative of the intensity profile.

Such a relation can be written in the form
\begin{equation} \label{eq:WFA_Q}
Q(\lambda)\approx-\gamma^2\lambda_0^4\,\cos2\Phi_B\,
	\frac{\d^2}{\d\lambda^2}I(\lambda)\;,
\end{equation}
where $\gamma$ is directly proportional to the \emph{squared} transverse component of the
magnetic field, and $\Phi_B$ is the direction of the magnetic field
projected on the plane-of-sky (POS). Without loss of generality, we can
always choose the reference direction for linear polarization on the POS
such that Stokes $U$ vanishes, in which case we can simply drop the
azimuth dependence in the above equation for Stokes $Q$. The explicit 
expression of $\gamma$ can again be expressed in terms of the Larmor 
frequency of the applied field, according to
\begin{equation} \label{eq:gamma2}
\gamma^2=\frac{1}{4}\,\bar{G}\,\frac{\nu_B^2}{c^2}\sin^2\Theta_B\;,
\end{equation}
where $\bar{G}$ is a modified effective Land\'e factor for linear
polarization \citep{LL04}.

Working analogously to the derivation of Eq.~(\ref{eq:wfa-moment_V}), and assuming $\Phi_B=0$, we find
\begin{eqnarray} \label{eq:wfa-moment_Q}
\calq_2&\approx&
-\gamma^2\lambda_0^4\
\int_{\lambda_a}^{\lambda_b}
	\d\lambda\;(\lambda-\lambda_0)^2\,\frac{\d^2}{\d\lambda^2}I(\lambda) 
=-\gamma^2\lambda_0^4\biggl[
	(\lambda-\lambda_0)^2\,\frac{\d}{\d\lambda}I(\lambda)\Bigr|_{\lambda_a}^{\lambda_b}
    -2
	(\lambda-\lambda_0)\,I(\lambda)\Bigr|_{\lambda_a}^{\lambda_b}
+2\int_{\lambda_a}^{\lambda_b}
\d\lambda\;I(\lambda)\biggr]
\nonumber \\
&\approx&-2\gamma^2\lambda_0^4\left[\cali_0
	-(\lambda_b-\lambda_a)\,I^\mathrm{b}\right]\;,
\end{eqnarray}
where, like in the case of Stokes $V$, we assumed that the spectral interval $(\lambda_a,\lambda_b)$ extends far out in the wings of the line where the intensity is dominated by the possible presence of a background signal $I^\mathrm{b}$, and therefore the spectral derivative of the intensity vanishes at the boundary of such an interval. Equation~(\ref{eq:wfa-moment_Q}) yields at once
\begin{equation} \label{eq:magn.diagn2}
\gamma^2\approx\frac{1}{2\lambda_0^4}\,
\frac{\calq_2}{\cali_0^\mathrm{b}-\cali_0}\;.
\end{equation}

It is important to remark that the inversion of Eq.~(\ref{eq:wfa-moment_Q}) directly in terms of $\gamma$ would require enforcing the positivity of the ratio $\calq_2/(\cali_0^\mathrm{b}-\cali_0)$ in Eq.~(\ref{eq:magn.diagn2}), even if a large signal noise could cause the measured value of that ratio to be negative. Such a rectification would bias the resulting error distribution away from the Gaussian model assumed for our linearized error-propagation framework. Thus, in order to preserve an unbiased error distribution and the applicability of our linearized error-propagation model, we chose to invert Eq.~(\ref{eq:wfa-moment_Q}) directly in terms of $\gamma^2$ instead of $\gamma$. This is the natural choice also because $\gamma^2$ is the physically meaningful quantity associated with the transverse Zeeman effect, which is quadratic in the magnetic field strength. We will later show how, for sufficiently large field strengths and photon counts, the ensuing positivity of the measured $\gamma^2$ value through Eq.~(\ref{eq:magn.diagn2}) allows us to derive an unbiased statistics for $\gamma$ itself. 

Similarly to the case of Stokes $V$, we can neglect any correlation of the background signal $\cali_0^\mathrm{b}$ with either $\cali_0$ or $\calq_2$.
Error propagation applied to Eq.~(\ref{eq:magn.diagn2}) as a function of the quantities $\calq_2$, $\cali_0$, and $\cali_0^\mathrm{b}$ then gives, analogously to the derivation of Eq.~(\ref{eq:sigmabet.tmp}),
\begin{eqnarray} \label{eq:sigmagam.tmp}
\sigma^2(\gamma^2)
&\approx&\frac{1}{4\lambda_0^8}\frac{\sigma^2(\calq_2)}{(\cali_0-\cali_0^\mathrm{b})^2}
	+\frac{1}{4\lambda_0^8}\,\calq_2^2\,
	\frac{\sigma^2(\cali_0)+\sigma^2(\cali_0^\mathrm{b})}{(\cali_0-\cali_0^\mathrm{b})^4}
    -\frac{1}{4\lambda_0^8}\,\calq_2\,
	\frac{2\,\mathrm{Cov(\calq_2,\cali_0)}}{(\cali_0-\cali_0^\mathrm{b})^3}
\nonumber \\
&\approx&\frac{1}{4\lambda_0^8}\frac{\sigma^2(\calq_2)}{(\cali_0-\cali_0^\mathrm{b})^2}
	+\gamma^4\,\frac{\cali_0+\cali_0^\mathrm{b}}{(\cali_0-\cali_0^\mathrm{b})^2} 
    +\frac{1}{\lambda_0^4}\,\gamma^2\,
	\frac{\mathrm{Cov(\calq_2,\cali_0)}}{(\cali_0-\cali_0^\mathrm{b})^2}\;.
\end{eqnarray}
Similarly to the case of Stokes $V$, we can approximate $\mathrm{Cov}(\calq_2,\cali_0)\approx\calq_2$ (see App.~\ref{app:covariances}) to finally arrive at the expression
\begin{equation} \label{eq:sigmagam}
\sigma^2(\gamma^2)
\approx\frac{r_Q^2}{\lambda_0^4\,(\cali_0-\cali_0^\mathrm{b})}\biggl(\frac{1}{4\lambda_0^4}\frac{\cali_4^\mathrm{l}+\cali_4^\mathrm{b}}{\cali_0-\cali_0^\mathrm{b}} 
-\frac{\gamma^4\lambda_0^4}{r_Q^2}\,\frac{\cali_0-3\cali_0^\mathrm{b}}{\cali_0-\cali_0^\mathrm{b}}\biggr)\;,
\end{equation}
having also recalled the derivation of the relation Eq.~(\ref{eq:v2mom}), to obtain (cf.\ Eq.~(\ref{eq:rSdef}))
\begin{equation} \label{eq:q2mom}
\sigma^2(\calq_2)
\approx\frac{\epsilon_I^2}{\epsilon_Q^2}\,\sigma^2(\cali_2)
\approx r_Q^2\,\cali_4\;.
\end{equation}
Similarly to the case of Fig.~\ref{fig:v1-s2}, the above relation can also be verified in the mean using Monte Carlo methods.

\begin{figure}[t!]
\centering
\includegraphics[width=.495\linewidth]{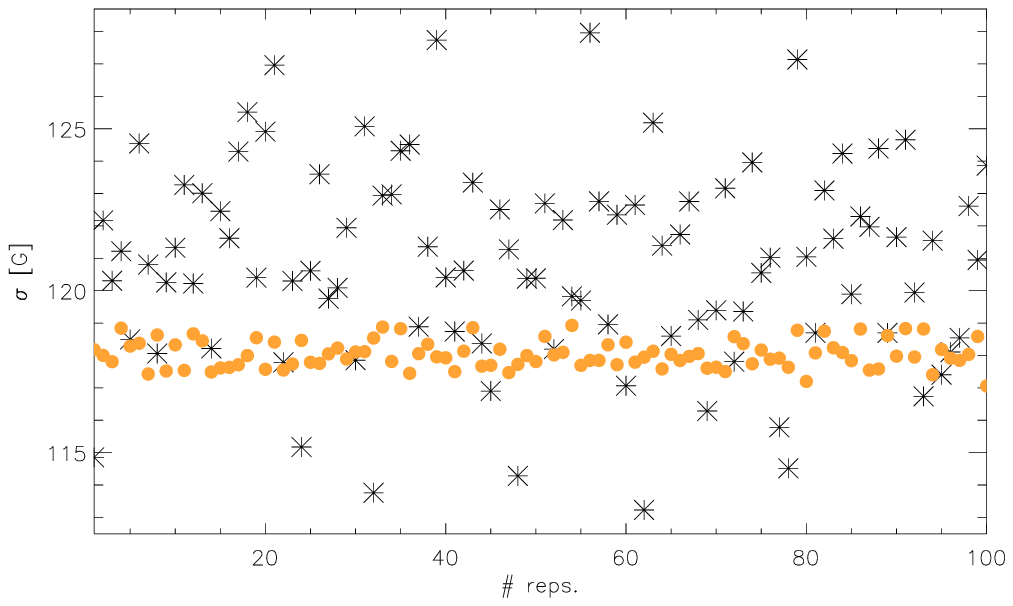}
\includegraphics[width=.495\linewidth]{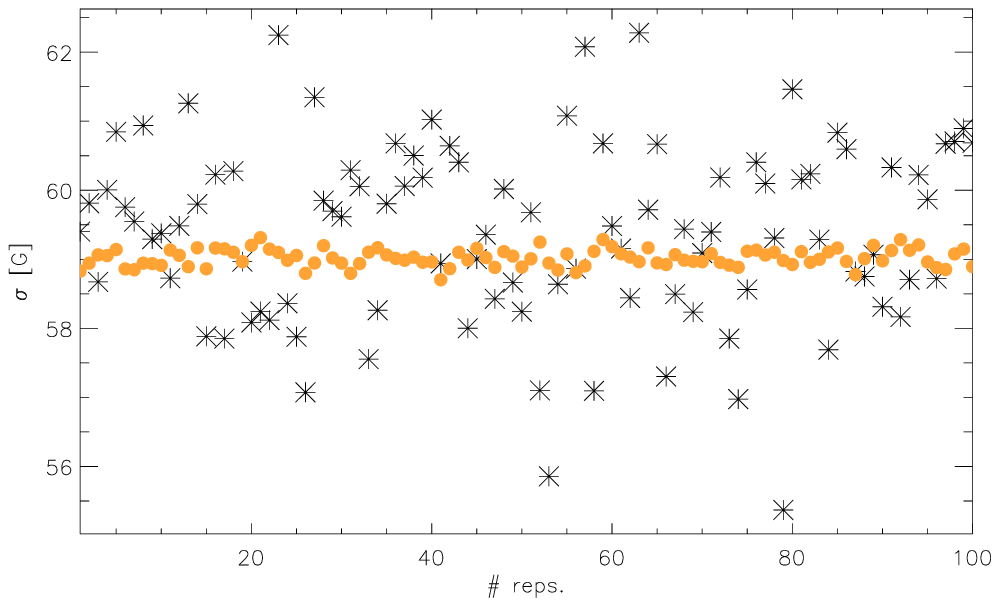}
\caption{\small \label{fig:sgmBmomentQ}
Numerical verification of the validity of the estimator (\ref{eq:sigmagam}) of the magnetic inference variance for a transverse magnetic field of 1000\,G. The testing conditions are similar to those of Fig.~\ref{fig:v1-s2}, with a background/signal ratio at line center of 10\% as measured at the detector and a spectral resolving power $R\approx20000$, assuming an ideal $(I,Q)$ spectro-polarimeter. The two panels show the statistical distributions of the absolute error on the inferred magnetic strength for two different integration times and corresponding SNR values at line center: $\rm SNR\approx 5000$ (left) and  $\rm SNR\approx 10000$ (right). For these values of the transverse field and SNR values, we can see the effect of the residual bias of the approximation (\ref{eq:gam2-gam}): for the case of $\mathrm{SNR}\approx5000$, the estimator gives an error of 118\,G compared to 121\,G from the Monte Carlo realizations, while for $\mathrm{SNR}\approx10000$ the estimation gives 59.0\,G compared to the Monte Carlo value of 59.4\,G.}
\end{figure}

Because the inference $\gamma^2$ of Eq.~(\ref{eq:magn.diagn2}) is unbiased and normally distributed over its full range, the error estimator (\ref{eq:sigmagam}) can be applied without restriction. Such an estimator can then be exploited to provide an error estimator of $\gamma$ itself. This is simply done by applying Eq.~(\ref{eq:EPvardiag}) to $\gamma=\sqrt{\gamma^2}$, 
\begin{equation} \label{eq:gam2-gam}
\sigma^2(\gamma)\approx \frac{\sigma^2(\gamma^2)}{4\gamma^2}\;,
\end{equation}
provided that $\sigma(\gamma^2)/\gamma^2\ll 1$ in order that the linearization assumption implied by Eq.~(\ref{eq:EPvardiag}) be applicable (see the concluding remark of Sect.~\ref{sec:errprop}). This last condition directly translates into a requirement of sufficiently strong fields and/or high SNR in order for the $\sigma(\gamma)$ estimator to remain unbiased. In App.~\ref{app:bias}, we derive in more detail the conditions that validate Eq.~(\ref{eq:gam2-gam}), and we evaluate the error bias on $\gamma$ when such conditions are only partially fulfilled.

Figure~\ref{fig:sgmBmomentQ} illustrates two examples of inference errors using Eqs.~(\ref{eq:sigmagam}) and (\ref{eq:gam2-gam}) in the case of a purely transverse field of 1000\,G, under observing conditions similar to the case of Fig.~\ref{fig:v1-s2} (see caption for details), but assuming much larger SNR values, which become necessary for magnetic diagnostics via the transverse Zeeman effect. Comparison of these results with the case of Stokes $V$ for the longitudinal Zeeman effect clearly demonstrates the penalization of the quadratic effect to infer the POS component of the field.
The two panels of Fig.~\ref{fig:sgmBmomentQ} show error distributions based on the variance estimator (\ref{eq:gam2-gam}). Similarly to the case of Stokes $V$, we find that the correction due to the covariance term $\mathrm{Cov}(\calq_2,\cali_0)$ in Eq.~(\ref{eq:sigmagam.tmp}) is completely masked by the spread of the estimator distributions shown in Fig.~\ref{fig:sgmBmomentQ}. Adding the next-order correction to Eq.~(\ref{eq:gam2-gam}) (see App.~\ref{app:bias}, in particular Eq.~(\ref{eq:last})) allows us to provide a less biased error estimate under more relaxed conditions than the one stated above.

As a last remark, we point out that spatial binning of the inferred $\gamma^2$ over $N$ pixels statistically reduces the $\sigma(\gamma^2)/\gamma^2$ error by $\sqrt N$ and the bias introduced by Eq.~(\ref{eq:gam2-gam}) by a factor $N$, consistently with the $\sqrt N$ increase of the SNR of binned signals. This is possible precisely because the variance estimator (\ref{eq:magn.diagn2}) is unbiased, unlike its square root, hence Eq.~(\ref{eq:gam2-gam}) should be applied only after binning. For a field that is not uniform over the binning area, of course, the resulting $\gamma$ estimates the rms transverse field over the area.

Once $\sigma^2(\gamma)$ is retrieved using Eq.~(\ref{eq:gam2-gam}), Eqs.~(\ref{eq:sigmabeta}) and (\ref{eq:sigmagam}) can be adopted to derive an expression that estimates the error on the magnetic field inclination with the LOS,
noting that, from Eqs.~(\ref{eq:beta}) and (\ref{eq:gamma2}),
\begin{equation} \label{eq:tanTheta}
\tan\Theta_B=2\,\frac{\bar{g}}{\sqrt{\bar{G}}}\,\frac{\gamma}{\beta}\;.
\end{equation}
The determination of the inclination angle via \emph{both} Stokes $V$ and $Q$ through Eq.~(\ref{eq:tanTheta}) evidently requires that both signatures of the longitudinal and transverse Zeeman effect must be detectable with 
sufficient SNR, and in particular that $\sin\Theta_B,\cos\Theta_B\ne0$. In such a 
case, using Eqs.~(\ref{eq:EPvar}) and (\ref{eq:EPrelgen}), the error propagation applied to the $\arctan$ of the expression on the RHS of Eq.~(\ref{eq:tanTheta}) gives
\begin{equation} \label{eq:TB}
\sigma^2(\Theta_B)
\approx \frac{\tan^2\Theta_B}{(1+\tan^2\Theta_B)^2}\!
\left[
\frac{\sigma^2(\beta)}{\beta^2}+\frac{\sigma^2(\gamma)}{\gamma^2}-2\,\frac{\mathrm{Cov}(\beta,\gamma)}{\beta\gamma}
\right].
\end{equation}

Evidently, the simultaneous measurement of Stokes $V$ and $Q$ generally requires the use of a polarimeter that adopts a full-Stokes modulation scheme specified by an appropriate modulation matrix $\mathbf{O}$. This leads to the question of the covariance $\mathrm{Cov}(\calv_1,\calq_2)$ between the two polarization moments entering Eq.~(\ref{eq:TB}). In App.~\ref{app:covariances} we show that such a covariance vanishes in the case of a dual-beam polarimeter with perfectly balanced beams, if the single-beam modulation matrix is characterized by a diagonal $\bm{\lambda}$ matrix (see App.~\ref{app:covariances} for the definition of $\bm{\lambda}$). In such a case, we simply have
%
\begin{equation} \label{eq:cov-moment}
2\,\frac{\mathrm{Cov}(\beta,\gamma)}{\beta\gamma}
=-\frac{\cali_0-3\cali_0^\mathrm{b}}{(\cali_0-\cali_0^\mathrm{b})^2}\;,
\end{equation}
which is seen to exactly cancel out the second addendum in $\sigma^2(\beta)/\beta^2$ (cf.\ Eq.~(\ref{eq:sigmabeta})), leaving a corresponding contribution from $\sigma^2(\gamma)/\gamma^2$ (cf.\ Eqs.~(\ref{eq:sigmagam}) and (\ref{eq:gam2-gam})) that amounts to $1/4$ of the expression (\ref{eq:cov-moment}). For a general polarimeter design, such covariance term may not be negligible, and in fact it may compete in magnitude with the contribution of Eq.~(\ref{eq:cov-moment}).
For the sake of brevity, we are not going to demonstrate numerically the error $\sigma^2(\Theta_B)$ that ensues from the integral-moment approach, leaving a more detailed analysis to the spectrally resolved case of the next section.

To conclude this section, we want to relate the error on the moment $\cals_m$ with the
signal-to-noise ratio (SNR) of an observation, which is ultimately the
quantifier of the sensitivity of the measurements and how this impacts
the spectro-polarimetric noise. Going back to
Eq.~(\ref{eq:noise}), the following inequality holds
\begin{eqnarray*}
\sigma^2(\cals_m)
&\approx&\sum_{i=1}^N
	(\lambda_i-\lambda_0)^{2m}\,\sigma_S^2(\lambda_i) 
\lesssim\sigma_S^2\sum_{i=1}^N
	(\lambda_i-\lambda_0)^{2m} 
\approx\frac{1}{\delta\lambda}\,\sigma_S^2\int_{\lambda_a}^{\lambda_b}\d\lambda\;
	(\lambda-\lambda_0)^{2m} \\
&\lesssim&\frac{\sigma_S^2}{2m+1}\,
	\frac{(\lambda_b-\lambda_0)^{2m+1}-(\lambda_a-\lambda_0)^{2m+1}}%
	{\delta\lambda}\;,
\end{eqnarray*}
where $\sigma_S$ is the largest noise expected across the line.
If we choose the bandwidth such that
$\lambda_{a,b}=\lambda_0\mp\Delta\lambda/2$, together with
Eq.~(\ref{eq:sampling}), we obtain
\begin{eqnarray}
\sigma^2(\cals_m)
&\lesssim&
	\frac{\sigma_S^2}{2m+1}
	\frac{2\,\Delta\lambda^{2m+1}}{2^{2m+1}}
\approx \frac{\sigma_S^2}{2m+1}
	\frac{(N\,\delta\lambda)^{2m+1}}{2^{2m}\delta\lambda} \nonumber \\
&\approx&
\frac{\delta\lambda^{2m}\,N^{2m+1}}{2^{2m}(2m+1)}\,\sigma_S^2\;,
\end{eqnarray}
that is,
\begin{equation}
\sigma(\cals_m) \lesssim
\frac{\delta\lambda^m}{2^m}\,\sqrt{\frac{N^{2m+1}}{2m+1}}\,\sigma_S\;.
\end{equation}

\section{Spectrally resolved analysis of Stokes signals}
\label{sec:deriv}

When polarimetric observations can achieve both high spectral resolution and high SNR, it is common to approach the WFA by dealing directly with the spectral derivatives of the Stokes profiles, Eqs.~(\ref{eq:WFA_V}) and (\ref{eq:WFA_Q}). In this case, one must perform the error propagation 
analysis of the signal directly through the spectral derivatives of the intensity.

In order to do this, for both cases of Stokes $V$ and $Q$, we can 
numerically approximate the spectral derivative of Stokes $I$, using 
measurements separated by the spectral sampling interval $\delta\lambda$. In particular, for the 1st-order derivative, we adopt the centered finite-difference scheme.
We thus have
\begin{eqnarray} \label{eq:derI}
J_\lambda&\equiv&\frac{\d}{\d\lambda}I_\lambda
\approx\frac{I_{\lambda+\delta\lambda}-I_{\lambda-\delta\lambda}}{2\,\delta\lambda} 
\approx \frac{R}{\lambda_0}
	(I_{\lambda+\delta\lambda}-I_{\lambda-\delta\lambda})\;,
\end{eqnarray}
which also allows us to estimate the noise on the derivative,
\begin{eqnarray} \label{eq:sgmId}
\sigma^2_J(\lambda)
&\approx& \frac{\sigma^2_I(\lambda+\delta\lambda)+\sigma^2_I(\lambda-\delta\lambda)}{4\,\delta\lambda^2} 
\approx \frac{I_{\lambda+\delta\lambda}+I_{\lambda-\delta\lambda}}{4\,\delta\lambda^2}\;.
\end{eqnarray}
%
%
%
The expression for the 2nd-order derivative is instead given by
\begin{eqnarray} \label{eq:der2I}
K_\lambda&\equiv&\frac{\d^2}{\d\lambda^2}I_\lambda
\approx\frac{I_{\lambda+\delta\lambda}-2I_\lambda+I_{\lambda-\delta\lambda}}{\delta\lambda^2} 
\approx\frac{4R^2}{\lambda_0^2}\left(I_{\lambda+\delta\lambda}-2I_\lambda+I_{\lambda-\delta\lambda}\right)\;,
\end{eqnarray}
and error propagation gives at once
\begin{eqnarray} \label{eq:sgmId2}
\sigma^2_K(\lambda)
&\approx&\frac{\sigma^2_I(\lambda+\delta\lambda)+4\sigma^2_I(\lambda)+\sigma^2_I(\lambda-\delta\lambda)}{\delta\lambda^4} 
 \approx
\frac{I_{\lambda+\delta\lambda}+4I_\lambda+I_{\lambda-\delta\lambda}}{\delta\lambda^4}\;.
\end{eqnarray}
%
%
%

 Similarly to the development of the previous section, we can invert Eqs.~(\ref{eq:WFA_V}) and (\ref{eq:WFA_Q}), respectively, to derive $\beta$ and $\gamma^2$ from the measured signals, and their errors driven by the SNR of the observations:
\begin{eqnarray}
\label{eq:Vder}
\beta&\approx&-\frac{1}{\lambda_0^2}\,\frac{V_\lambda}{J_\lambda}\;, \\
\label{eq:Qder}
\gamma^2&\approx&-\frac{1}{\lambda_0^4}\,
\frac{Q_\lambda}{K_\lambda}\;.
\end{eqnarray}
In particular, Eqs.~(\ref{eq:Vder}) and (\ref{eq:Qder}) should be compared respectively with 
Eqs.~(\ref{eq:magn.diagn}) and (\ref{eq:magn.diagn2}). 

Similarly to what we already discussed for the derivation of
Eq.~(\ref{eq:v2mom}), the noise on Stokes $V$ and $Q$ will depend on the
specific demodulation scheme. Thus, if we indicate with $S$ any of the
Stokes parameters, we simply have 
\begin{equation} \label{eq:Stokes_noise}
\sigma^2_S(\lambda)
\approx\frac{\epsilon_I^2}{\epsilon_S^2}\,\sigma^2_I(\lambda)
\approx r_S^2\,I_\lambda\;,
\end{equation}
having assumed again Poissonian noise.

It is important to remark that Eqs.~(\ref{eq:sgmId}), (\ref{eq:sgmId2}), and
(\ref{eq:Stokes_noise}) remain true in the presence of a background continuum, 
and therefore the following development applies also to such a case for both emission and absorption lines.
By applying Eq.~(\ref{eq:EPreldiag}) to
Eq.~(\ref{eq:Vder}), and recalling Eqs.~(\ref{eq:sgmId}) and (\ref{eq:Stokes_noise}), 
we have
\begin{eqnarray*}
\frac{\sigma^2(\beta)}{\beta^2}
&\approx& \frac{\sigma^2_V(\lambda)}{V^2_\lambda} 
+	\frac{\sigma^2_J(\lambda)}{J^2_\lambda} 
\approx \frac{1}{\beta^2\lambda_0^4}\frac{\sigma^2_V(\lambda)}{J^2_\lambda} 
+ 	\frac{\sigma^2_J(\lambda)}{J^2_\lambda} \nonumber \\
&\approx& \frac{r_V^2}{\beta^2\lambda_0^4}\frac{I_\lambda}{J^2_\lambda} 
+ 	\frac{I_{\lambda+\delta\lambda}+I_{\lambda-\delta\lambda}}{4\,\delta\lambda^2 J^2_\lambda} \nonumber \\
&\approx&\frac{I_\lambda}%
	    {4\,\delta\lambda^2 J^2_\lambda}
\left(\frac{4\,r_V^2\,\delta\lambda^2}{\beta^2\lambda_0^4}+
\frac{I_{\lambda+\delta\lambda}+I_{\lambda-\delta\lambda}}{I_\lambda}
\right)\;.
\end{eqnarray*}
or
\begin{equation} \label{eq:sigma2beta}
\sigma^2(\beta)
\approx\frac{1}{\lambda_0^2}\frac{I_\lambda}%
	    {(I_{\lambda+\delta\lambda}-I_{\lambda-\delta\lambda})^2} 
\left(\frac{4\,r_V^2\,\delta\lambda^2}{\lambda_0^2}+\beta^2\lambda_0^2\,
\frac{I_{\lambda+\delta\lambda}+I_{\lambda-\delta\lambda}}{I_\lambda}
\right).
\end{equation}
The term with $\mathrm{Cov}(V_\lambda,J_\lambda)$ in this case vanishes identically, because by adopting the ``centered'' form (\ref{eq:derI}) of the $J_\lambda$ derivative the two observables do not share sampled signals. A direct way to see this is to carry out the error propagation by rewriting Eq.~(\ref{eq:Vder}) explicitly in terms of the six \emph{independently 
measured} signals $v^\pm_\lambda$ and
$v^\pm_{\lambda\pm\delta\lambda}$ acquired by an ideal $(I,V)$ dual-beam polarimeter,
\begin{displaymath}
\beta=-\frac{2\,\delta\lambda}{\lambda_0^2}\,\frac{v^+_\lambda-v^-_\lambda}%
{v^+_{\lambda+\delta\lambda}+v^-_{\lambda+\delta\lambda}-
 v^+_{\lambda-\delta\lambda}-v^-_{\lambda-\delta\lambda}}\;,
\end{displaymath}
and verify that we arrive at an expression of the relative noise identical to Eq.~(\ref{eq:sigma2beta}).
We note that, for sufficiently dense spectral samplings of the profile,
it is possible to additionally approximate 
$2I_\lambda\approx I_{\lambda+\delta\lambda}+I_{\lambda-\delta\lambda}$, leading to a
further simplification of this expression.

%

%
%

Similarly, in the case of Stokes $Q$, \emph{if we explicitly neglect the contribution from} $\mathrm{Cov}(Q_\lambda,K_\lambda)$ we find
\begin{eqnarray*}
\frac{\sigma^2(\gamma^2)}{\gamma^4}
&\approx& \frac{\sigma^2_Q(\lambda)}{Q^2_\lambda} 
+	  \frac{\sigma^2_K(\lambda)}{K^2_\lambda} 
\approx 
\frac{1}{\gamma^4\lambda_0^8}\,
	\frac{\sigma^2_Q(\lambda)}{K^2_\lambda} 
+ \frac{\sigma^2_K(\lambda)}{K^2_\lambda} \nonumber \\ 
&\approx&
\frac{r_Q^2}{\gamma^4\lambda_0^8}
	\frac{I_\lambda}{K^2_\lambda} 
+ 
\frac{I_{\lambda+\delta\lambda}+4I_\lambda+I_{\lambda-\delta\lambda}}{\delta\lambda^4
K^2_\lambda} \nonumber \\
&\approx& 
 	\frac{I_\lambda}%
	    {\delta\lambda^4
K^2_\lambda} 
\left(
\frac{r_Q^2\,\delta\lambda^4}{\gamma^4\lambda_0^8}+
\frac{I_{\lambda+\delta\lambda}+4I_\lambda+I_{\lambda-\delta\lambda}}{I_\lambda}
\right),
\end{eqnarray*}
or
\begin{eqnarray*}
\sigma^2(\gamma^2)
&\approx& \frac{1}{\lambda_0^4}
 	\frac{I_\lambda}%
	    {(I_{\lambda+\delta\lambda}-2I_\lambda+I_{\lambda-\delta\lambda})^2} 
\left(
\frac{r_Q^2\,\delta\lambda^4}{\lambda_0^4}
+\gamma^4\lambda_0^4\,
\frac{I_{\lambda+\delta\lambda}+4I_\lambda+I_{\lambda-\delta\lambda}}{I_\lambda}
\right).
\end{eqnarray*}
However, analogously to the case of $\beta$, we can estimate the impact of the neglected covariance term by assuming the case of an ideal $(I,Q)$ dual-beam polarimeter, and rewriting the expression of $\gamma^2$ in terms of the component signals,
\begin{displaymath}
\gamma^2=-\frac{\delta\lambda^2}{\lambda_0^4}\,\frac{q^+_\lambda-q^-_\lambda}%
{q^+_{\lambda+\delta\lambda}+q^-_{\lambda+\delta\lambda}-
 2(q^+_\lambda+q^-_\lambda)
 +q^+_{\lambda-\delta\lambda}+q^-_{\lambda-\delta\lambda}}\;.
\end{displaymath}
This time the error propagation applied to the six signals $q^\pm_\lambda$ and
$q^\pm_{\lambda\pm\delta\lambda}$ leads to the appearance of a correction to the expression of $\sigma^2(\gamma^2)$ above, given by
\begin{displaymath}
\delta\sigma^2(\gamma^2)\approx\frac{4\gamma^4}%
{I_{\lambda+\delta\lambda}-2I_\lambda+I_{\lambda-\delta\lambda}}\;,
\end{displaymath}
and therefore, after some simple algebra, we obtain
\begin{eqnarray} \label{eq:sigma2gamma}
\sigma^2(\gamma^2)
&\approx& \frac{1}{\lambda_0^4}
 	\frac{I_\lambda}%
{(I_{\lambda+\delta\lambda}-2I_\lambda+I_{\lambda-\delta\lambda})^2} 
\left(
\frac{r_Q^2\,\delta\lambda^4}{\lambda_0^4}
+\gamma^4\lambda_0^4\,
\frac{5I_{\lambda+\delta\lambda}-4I_\lambda+5I_{\lambda-\delta\lambda}}{I_\lambda}
\right).
\end{eqnarray}
%

We note how both Eqs.~(\ref{eq:sigma2beta}) and (\ref{eq:sigma2gamma})
show that the noise on the respective magnetic field projections 
decreases inversely proportionally with the central wavelength of 
the spectral line, as expected in the case of the Zeeman effect (see
discussion after Eq.~(\ref{eq:error-beta})).
We also observe that the two equations have a similar structure, 
sharing in particular the same ratio involving the intensity signals 
around the wavelength $\lambda$ of observation. In the presence of 
a constant background $I^\mathrm{b}$, the denominator of that ratio 
is unchanged, whereas the numerator acquires a contribution 
$\delta\lambda\,I^\mathrm{b}$ that adds to the overall noise.

Similarly to the case of the integral-moment approach, under the proper conditions of magnetic field strength and SNR that make the error statistics on $\gamma$ unbiased (see discussion about Eq.~(\ref{eq:gam2-gam})), Eqs.~(\ref{eq:sigma2beta}) and (\ref{eq:sigma2gamma}) can be used to estimate the error on the magnetic geometry via Eq.~(\ref{eq:TB}), also in the case of spectrally resolved measurements. The covariances between the derived signals $V_\lambda$ and $Q_\lambda$ and the first-order derivative $J_\lambda$ are both zero because they do not share sampled signals, when we adopt the ``centered'' form of $J_\lambda$. On the other hand, the covariances with the second-order $K_\lambda$ are non vanishing, and satisfy the relation
\begin{equation} \label{eq:SKcov}
\frac{\mathrm{Cov}(V_\lambda,K_\lambda)}{V_\lambda\,K_\lambda}=\frac{\mathrm{Cov}(Q_\lambda,K_\lambda)}{Q_\lambda\,K_\lambda}=-\frac{2}{\delta\lambda^2 K_\lambda}\approx
-\frac{2}{I_{\lambda+\delta\lambda}-2I_\lambda+I_{\lambda-\delta\lambda}}\;.
\end{equation}
Through a similar argument to that leading to Eq.~(\ref{eq:cov-moment}), we also neglect the covariance between $V_\lambda$ and $Q_\lambda$ during the measurement process, assuming the ideal case of a perfectly balanced dual-beam polarimeter designed around a modulation scheme with a diagonal $\bm{\lambda}$ matrix (see App.~\ref{app:covariances}).
We thus find, using Eqs.~(\ref{eq:Vder}),(\ref{eq:Qder}), and (\ref{eq:SKcov}),
\begin{eqnarray} \label{eq:TB1}
2\,\frac{\mathrm{Cov}(\beta,\gamma)}{\beta\gamma} &\approx& 
\frac{\mathrm{Cov}(J_\lambda,K_\lambda)}{J_\lambda K_\lambda} -
\frac{\mathrm{Cov}(V_\lambda,K_\lambda)}{V_\lambda K_\lambda} 
\approx
\frac{\sigma^2_I(\lambda+\delta\lambda)-\sigma^2_I(\lambda-\delta\lambda)}{2\,\delta\lambda^3 J_\lambda\,K_\lambda}
+\frac{2}{\delta\lambda^2 K_\lambda} \nonumber \\
&\approx&
\frac{I_{\lambda+\delta\lambda}-I_{\lambda-\delta\lambda}}{2\,\delta\lambda^3 J_\lambda\,K_\lambda}
+\frac{2}{\delta\lambda^2 K_\lambda} 
\approx \frac{1}{\delta\lambda^2 K_\lambda}
+\frac{2}{\delta\lambda^2 K_\lambda}\approx
\frac{3}{I_{\lambda+\delta\lambda}-2I_\lambda+I_{\lambda-\delta\lambda}}\;.
\end{eqnarray}
This result, along with Eqs.~(\ref{eq:sigma2beta}) and (\ref{eq:sigma2gamma}), allows us to cast the variance of $\Theta_B$ explicitly as a function of the strength and geometry of the magnetic field. This is done by recalling the derivations of Eqs.~(\ref{eq:sigma2beta}) and (\ref{eq:sigma2gamma}), through which we rewrite, also recalling Eq.~(\ref{eq:gam2-gam}),
\begin{eqnarray} \label{eq:TB2}
\frac{\sigma^2(\beta)}{\beta^2}+\frac{\sigma^2(\gamma)}{\gamma^2}
&\approx& 
\frac{\sigma^2(\beta)}{\beta^2}+\frac{1}{4}\frac{\sigma^2(\gamma^2)}{\gamma^4} \nonumber \\
&\approx& 
\frac{4\,r_V^2\,\delta\lambda^2}{\beta^2\lambda_0^4}\frac{I_\lambda}{(I_{\lambda+\delta\lambda}-I_{\lambda-\delta\lambda})^2} 
+ 	\frac{I_{\lambda+\delta\lambda}+I_{\lambda-\delta\lambda}}{(I_{\lambda+\delta\lambda}-I_{\lambda-\delta\lambda})^2} \nonumber \\
&&{}+\frac{r_Q^2\,\delta\lambda^4}{\gamma^4\lambda_0^8}
	\frac{I_\lambda}{4\,(I_{\lambda+\delta\lambda}-2I_\lambda+I_{\lambda-\delta\lambda})^2} 
+\frac{5I_{\lambda+\delta\lambda}-4 I_\lambda+5I_{\lambda-\delta\lambda}}{4\,(I_{\lambda+\delta\lambda}-2I_\lambda+I_{\lambda-\delta\lambda})^2}\;
\end{eqnarray}
By using Eqs.~(\ref{eq:TB1}) and (\ref{eq:TB2}), after much tedious algebra, the variance $\sigma^2(\Theta_B)$ of Eq.~(\ref{eq:TB}) can be expressed in the form
\begin{eqnarray}  \label{eq:sigma2Theta}
\sigma^2(\Theta_B)
&\approx& \sin^2\Theta_B\cos^2\Theta_B\, 
\bigl[f_1^2(\Theta_B)\,\rho_\lambda+f_2^2(\Theta_B)\,\tau_\lambda+\psi_\lambda\bigr]\;;
\end{eqnarray}
here
%
\begin{subequations} \label{eq:ffun}
\begin{eqnarray}
f_1(\alpha)
&=&\frac{r_V\,\delta\lambda/\lambda_0}{\bar{g}\,\kappa
\lambda_0 B\cos\alpha}\;, \\
f_2(\alpha)
&=&2\,\frac{r_Q\,\delta\lambda^2/\lambda_0^2}{\bar{G}\,\kappa^2\lambda_0^2
B^2\sin^2\alpha}\;,
\end{eqnarray}
\end{subequations}
with $\kappa\approx 4.6686\times10^{-12}\,\rm nm^{-1}G^{-1}$, and
\begin{subequations} \label{eq:profiles}
\begin{eqnarray}
\rho_\lambda&=&\frac{4I_\lambda}{(I_{\lambda+\delta\lambda}-I_{\lambda-\delta\lambda})^2}
\equiv\frac{I_\lambda}{\delta\lambda^2 J_\lambda^2}\;, \\
\label{eq:tau}
\tau_\lambda&=&\frac{I_\lambda}{(I_{\lambda+\delta\lambda}-2I_\lambda+I_{\lambda-\delta\lambda})^2}
\equiv\frac{I_\lambda}{\delta\lambda^4 K_\lambda^2}\;, \\
\label{eq:psi}
\psi_\lambda
&=&\frac{3}{2}\,\tau_\lambda
\left(2-\frac{I_{\lambda+\delta\lambda}+I_{\lambda-\delta\lambda}}{2I_\lambda}\right)
+\frac{(I_{\lambda+\delta\lambda}+I_{\lambda-\delta\lambda})K_\lambda-4 J_\lambda^2}
{4\,\delta\lambda^2 J^2_\lambda K_\lambda}\;.
\end{eqnarray}
\end{subequations}
%
%
For increasingly dense spectral samplings (i.e., decreasing $\delta\lambda$), the first addendum in the expression of $\psi_\lambda$ dominates quadratically over the second addendum, while $I_{\lambda+\delta\lambda}+I_{\lambda-\delta\lambda}\to 2I_\lambda$, thus
\begin{displaymath}
\psi_\lambda
\underset{\delta\lambda\to 0}{\approx}\frac{3}{2}\,\tau_\lambda\;.
\end{displaymath}
Because $I_\lambda\sim\delta\lambda$ \emph{for a fixed integration time}, Eq.~(\ref{eq:tau}) shows that the \emph{magnetic independent} contribution from $\psi_\lambda$ to Eq.~(\ref{eq:sigma2Theta}) grows like $\delta\lambda^{-5}$ by increasing the spectral sampling/resolution. In contrast, Eqs.~(\ref{eq:ffun}) show that the \emph{magnetic dependent} contributions to the same error only grow like $\delta\lambda^{-1}$. It is worth noting that $\psi_\lambda\sim\sigma^2_K(\lambda)/(4\,K^2_\lambda)$ for $\delta\lambda\to 0$ (cf.\ Eq.~(\ref{eq:sgmId2})), which is exactly the contribution of the noise to the second derivative $K_\lambda$ to $\sigma^2(\gamma)/\gamma^2$. This clarifies that the divergence of the variance (\ref{eq:sigma2Theta}) for $\delta\lambda\to 0$ is a direct consequence of the ill-posedness of the numerical differentiation of noised data. A regularization of these derivatives would require the smoothing of the noised profiles over a finite spectral scale that is kept fixed as  $\delta\lambda\to 0$, at the cost of spectral resolution. This is in fact what the moment approach to magnetic inference discussed in the previous section implicitly does through the integration by parts in Eqs.~(\ref{eq:wfa-moment_V}) and (\ref{eq:wfa-moment_Q}), leading in that case to the variance (\ref{eq:TB}), which is instead immune to the breakdown encountered in the derivative approach.

\begin{figure}[t!]
\centering
\includegraphics[height=2.85in]{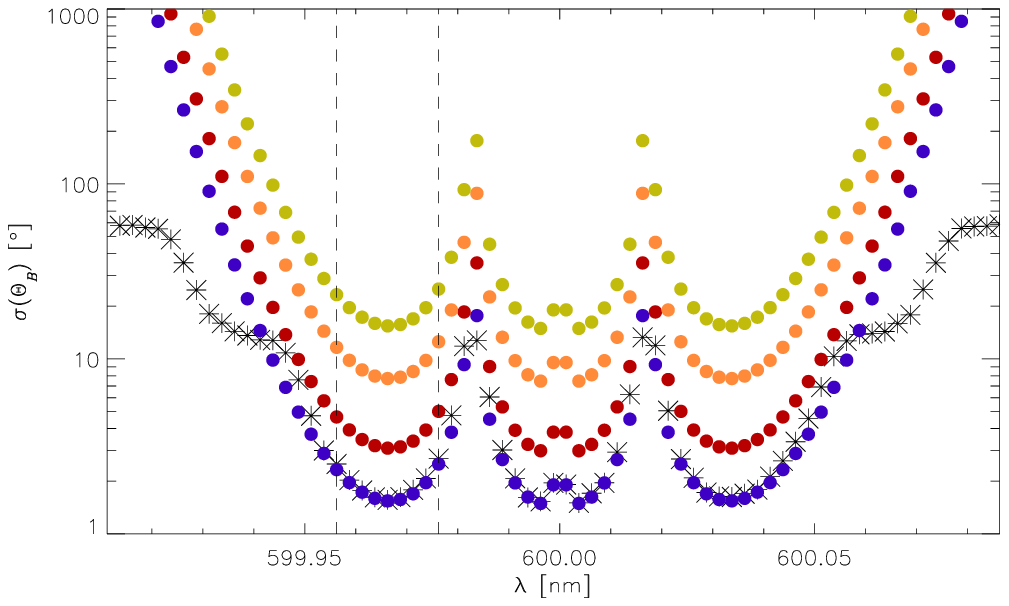}
\includegraphics[height=2.85in]{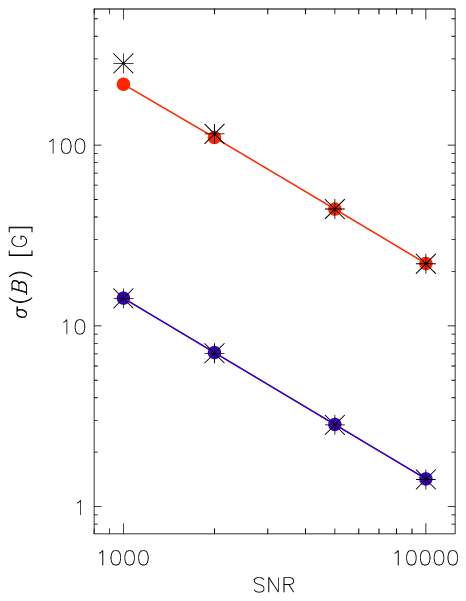}
\caption{\emph{Left:} plots of the error $\sigma(\Theta_B)$ of Eq.~(\ref{eq:sigma2Theta}) as a function of wavelength across a spectral line (dot symbols), for a 1000\,G field inclined by $\Theta_B=60^\circ$ from the LOS. For this example, we assumed a spectral resolving power $R\approx 120000$ and approximate SNR values of 10000 (blue dots), 5000 (red dots), 2000 (orange dots), and 1000 (green dots) at line center. Like in the previous examples, a background of 10\% relative to the peak signal is assumed also in this model. For the case of $\rm SNR \approx 10000$, the star symbols represent the Monte Carlo standard deviation of $\Theta_B$ from computing Eq.~(\ref{eq:tanTheta}) over 1000 random realizations of the Poissonian noise of the signals used for the calculation of the wavelength dependent values of $\beta$ and $\gamma$ via Eqs.~(\ref{eq:Vder}) and (\ref{eq:Qder}). For simplicity, we assumed $\bar{g}=\bar{G}=1$, and the ideal case of an optimally efficient full-Stokes polarimeter with $r_Q=r_V=\sqrt{3}$. We note the almost perfect overlap of the statistical and estimated errors as a function of wavelength in the high SNR regime well into the line wings. The vertical dashed lines identify the spectral range in the wings of the line where the error estimation is minimum, encompassing in this example 9 spectral points.
\emph{Right:} plots of the estimated errors on the field projections $B_{\rm LOS}$ (blue) and $B_{\rm POS}$ (red) as a function of the SNR, after averaging over the 9 spectral points contained by the spectral window in the left panel. The crosses correspond to the actual errors computed from the Monte Carlo realizations.}
\label{fig:sgmTB}
\end{figure}

In general, we find that the magnetic dependent error on the inference of $\Theta_B$ is largely dominated by the $f_2(\Theta_B)$ contribution. Evidently, for weaker and/or mostly longitudinal fields, this contribution can rapidly blow up the error budget on $\Theta_B$ via the quadratic denominator of $f_2(\Theta_B)$ (see Eqs.~(\ref{eq:ffun})). From the argument presented in the previous paragraph we can also conclude that this error is critically sensitive to both the SNR and the spectral resolution of the observation.

To demonstrate this last point, in the left panel of Fig.~\ref{fig:sgmTB} we plot different realizations of the error $\sigma(\Theta_B)$ as a function of wavelength across the spectral line domain (dot symbols), assuming a field of 1000\,G with an inclination of $60^\circ$ from the LOS, a spectral resolving power $R\approx 120000$, and SNR values of 10000 (blue dots), 5000 (red dots), 2000 (orange dots), and 1000 (green dots). These plots demonstrate that the error estimator (\ref{eq:sigma2Theta}) provides a reliable approximation of the variance of actual random realizations of $\Theta_B$ (star symbols, shown here for the case of $\rm SNR\approx 10000$) inside a spectral band (e.g., the one delimited by the vertical dashed lines) broadly within the spectral range of applicability of the WFA expressions (\ref{eq:WFA_V}) and (\ref{eq:WFA_Q}). The agreement between the spectral curves for the estimated and the statistical errors is found to improve rapidly for stronger fields and larger SNR values. This is clearly demonstrated by the largest SNR case in Fig.~\ref{fig:sgmTB}. For lower SNR values, this agreement deteriorates, the discrepancy being dominated by the larger contribution associated with the transverse component and its estimate $\sigma^2(\gamma)$ (cf.\ Eq.~(\ref{eq:sigma2gamma})). The example of Fig.~\ref{fig:sgmTB} of a 1000\,G field with a significant projection towards the observer ($B_\mathrm{LOS}=500$\,G in this case), detected with a SNR of at least ${\sim}\,10^3$, is representative of the minimum conditions that must be met for a reliable estimation of the field geometry.
We note that the statistical error on $\Theta_B$ flattens out in the far wings, where the line signal is zero and the error is completely dominated by the noise of the background signal. Therefore, as there is no inference accuracy left at those wavelengths, the estimated $\Theta_B$ is practically randomly distributed over the interval $(0^\circ,180^\circ)$, and the standard deviation saturates to a value close to that of a uniformly distributed random variable, i.e., to $\sim 90^\circ/\sqrt{3}\approx 52^\circ$.

In the right panel of Fig.~\ref{fig:sgmTB}, we also plot the inference errors on the field projections $B_{\rm LOS}$ (blue dots and curves) and $B_{\rm POS}$ (red dots and curves), for the same field strength and geometry considered in the left panel. These errors were estimated using Eqs.~(\ref{eq:sigma2beta}) and (\ref{eq:sigma2gamma}), along with Eq.~(\ref{eq:gam2-gam}). For these plots, the estimated errors were further averaged over the 9 wavelength points comprised by the spectral window in the left panel (see, e.g., Eq.~(\ref{eq:avgest2})). The crosses show the corresponding statistical errors from the Monte Carlo realizations. These curves confirm the expected 1/SNR improvement of the inference error for both field projections, and the significantly smaller (about 15$\times$) error for the longitudinal projection, because of the linear dependence of the Stokes $V$ signature on it. Similarly to the case of Fig.~\ref{fig:sgmBmomentQ}, in the worst case scenario of magnetic inference at lower SNR values, the error estimator on $B_{\rm POS}$ is negatively biased compared to the true statistical error of the Monte Carlo realizations. This is again the effect of the bias from the approximation (\ref{eq:gam2-gam}) already discussed in the context of Fig.~\ref{fig:sgmBmomentQ}.

We must remark that the spectral window adopted for the averaging of the magnetic projection errors in the right panel of Fig.~\ref{fig:sgmTB} corresponds to the one that minimizes $\sigma(\Theta_B)$, as stated in the caption. On the other hand, although this window may still lie within the spectral range of applicability of the WFA, it does not correspond necessarily to the spectral region where the $\sigma(B)$ error for either of the two field projections is also minimized (see, e.g., Fig.~\ref{fig:deriv_test} below, where an optimal window for the inference of $B_{\rm LOS}$ is instead identified).


\begin{figure*}[t!]
\centering
\includegraphics[width=.495\linewidth]{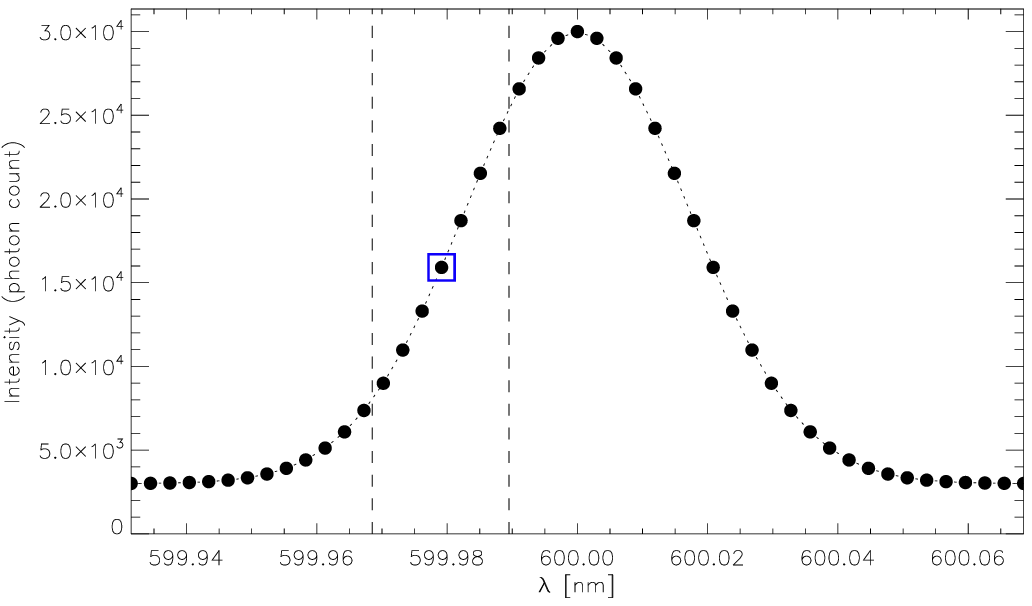}
\includegraphics[width=.495\linewidth]{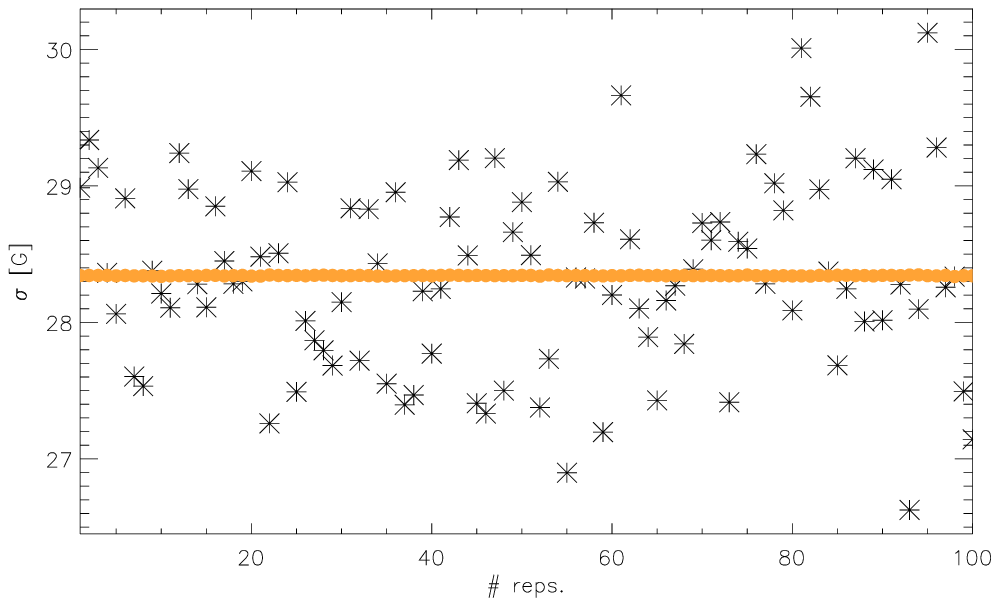}
\includegraphics[width=.495\linewidth]{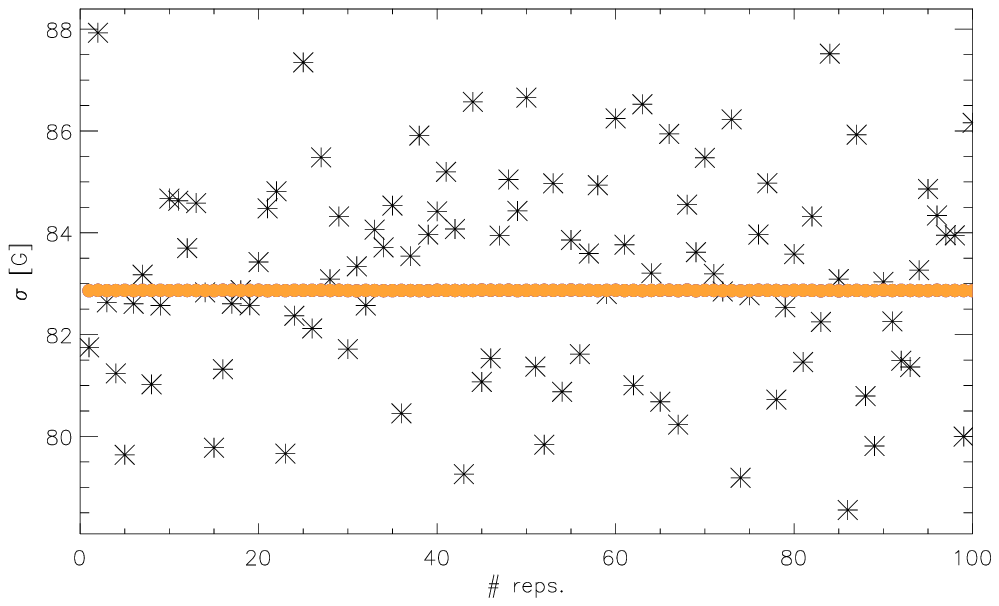}
\includegraphics[width=.495\linewidth]{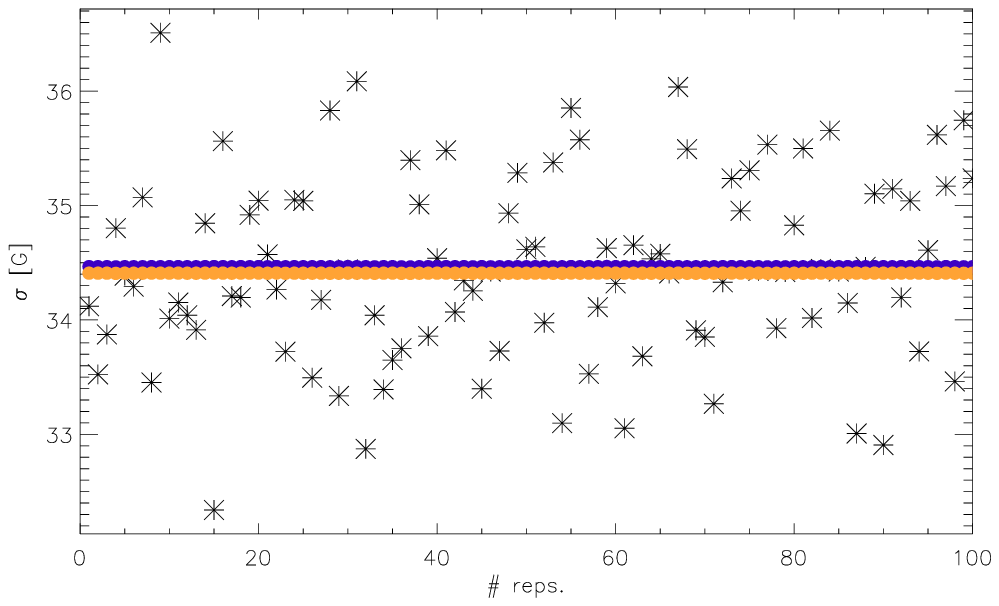}
\caption{\small \label{fig:deriv_test}
Performance comparison of the algebraic estimators of 
Eqs.~(\ref{eq:sigma2beta}), (\ref{eq:avgest2}), and (\ref{eq:meanest}) against the 
statistical realization of the inference error on the longitudinal 
magnetic strength via Eq.~(\ref{eq:Vder}) for a field of 100\,G along the LOS. 
The top-left panel shows the line profile adopted for this analysis,
simulating a condition of dense spectral sampling ($R\approx100000$), 
along with the bandpass used for the application of 
Eq.~(\ref{eq:meanest}), encompassing 7 spectral points; 
the central point inside the blue square identifies the wavelength 
for the application of the ``single-point'' estimator of 
Eq.~(\ref{eq:sigma2beta}).
The top-right panel shows the estimated error via the integral-moment
approach, similar to the plots shown in Fig.~\ref{fig:sgmBmoment}.
The two bottom panels show the distribution of the magnetic-inference error
and their algebraic estimators, in the case of the single-point inversion 
(bottom-left) and using all 7 points in the bandpass (bottom-right). (See detailed 
description in the text for the meaning of the three different symbols
and a discussion of the results.)}
\end{figure*}

To conclude this section, it is instructive to compare the expressions of the variance of Eq.~(\ref{eq:sigma2beta}) with the one we derived by the integral-moment approach, Eq.~(\ref{eq:error-beta}). In particular, in Eq.~(\ref{eq:sigma2beta}) 
it is notable the absence of an explicit  
line broadening term, which in Eq.~(\ref{eq:error-beta}) derives from the
evaluation of the 2nd-order moment of the intensity $\cali_2$ appearing
in Eq.~(\ref{eq:sigmabeta}). In order to reconcile this seeming 
discordance, we assume for simplicity the absence of a background 
signal (i.e., $\cali_m^{\rm b}=0$) in Eq.~(\ref{eq:sigmabeta}) and rewrite 
Eq.~(\ref{eq:sigma2beta}) in the simpler approximate form, also assuming for simplicity $r_V\approx 1$,
\begin{eqnarray} \label{eq:sgmB_simpl}
\sigma^2(\beta)
&\approx&\frac{1}{\lambda_0^2}\frac{I_\lambda}%
	    {\delta\lambda^2 J_\lambda^2}
\left(\frac{\delta\lambda^2}{\lambda_0^2}+\frac{1}{2}\,\beta^2\lambda_0^2\right) 
\approx\frac{1}{\lambda_0^2 I_\lambda}\frac{I^2_\lambda}%
	    {J_\lambda^2}
\left(\frac{1}{\lambda_0^2}+\frac{1}{2}\frac{\beta^2\lambda_0^2}{\delta\lambda^2}\right)\;.
\end{eqnarray}
By noting that at the inflection point of the Gaussian profile (see also Eq.~(\ref{eq:gaussian2moment}))
\begin{displaymath}
\frac{I^2_\lambda}{J_\lambda^2}
\approx \frac{\Delta^2_T+\Delta^2_\mathrm{PSF}}{2}
\approx \frac{\cali_2}{\cali_0}\;,
\end{displaymath}
we can rewrite Eq.~(\ref{eq:sgmB_simpl}) in a form closer to Eq.~(\ref{eq:sigmabeta}),
\begin{equation} \label{eq:sigma2betamod}
\sigma^2(\beta)
\approx\frac{1}{\lambda_0^2 I_\lambda}%
\left(\frac{1}{\lambda_0^2}\frac{\cali_2}{\cali_0}+\beta^2\lambda_0^2\,
\frac{\Delta^2_T+\Delta^2_\mathrm{PSF}}{4\,\delta\lambda^2}\right)\;.
\end{equation}
%
%
%
However, there remains a fundamental difference between the structures of
Eqs.~(\ref{eq:sigma2betamod}) and (\ref{eq:sigmabeta}), due to the
presence of the thermal Doppler width in the magnetic contribution to
the variance of Eq.~(\ref{eq:sigma2betamod}). This is not 
surprising, since in Eq.~(\ref{eq:sigma2betamod}) the magnetic inference 
is wavelength dependent, and specifically restricted to the spectral 
range around the inflection point of the intensity profile. While the 
estimates of $\beta$ correspond well between the two  
approaches, it can be expected that, in the spectral-derivative 
method, the variance should manifest a fundamentally different 
sensitivity to the magnetic strength and thermal width combined because of its wavelength dependence.

\section{Performance comparison of error estimators}
\label{sec:comparison}

We conclude this work by comparing the performance of the
magnetic inference errors based on the integral-moment and the spectral-derivative approaches of the previous sections.

The calculation of integral moments implies the combination of signals
acquired at different wavelength points, and we have found that the relative 
error is largely independent of the spectral resolution of the
observation, for a fixed integration time (i.e., preserving the line-integrated photon count), and as long as the line width is sampled finely enough 
(see right panel of Fig.~\ref{fig:test_resol}, and the discussion after Eq.~(\ref{eq:error-beta})). In the examples
used for this study, the main competing mechanism is the thermal 
broadening, because of the fundamental WFA condition. The error derived from the spectral-derivative approach, e.g.,
Eq.~(\ref{eq:sigma2beta}) for the longitudinal field strength $\beta$,
depends instead on the specific wavelength of the measurement.

In order to compare the two approaches, let us assume that we have
identified $n$ points across the spectrum where Eqs.~(\ref{eq:Vder}) 
and (\ref{eq:sigma2beta}) are applicable. These will belong to a spectral
interval $w$ where the spectral derivative $J_\lambda$ is \emph{maximum and
approximately constant} with a value that we indicate with $J_w$. The
value of $\beta$ will then be estimated as the mean of the $\beta_i$
values derived from Eq.~(\ref{eq:Vder}) at the points $\lambda_i$ within
$w$, for $i=1,\ldots,n$, i.e.,
\begin{eqnarray}
\label{eq:avgest1}
\bar\beta
&=&\frac{1}{n}\sum_{i=1}^n\beta_i\;, \\
\label{eq:avgest2}
\sigma^2(\bar\beta)
&\approx&\frac{1}{n^2}\sum_{i=1}^n\sigma^2(\beta_i)\;,
\end{eqnarray}
where the second equation gives the \emph{error on the mean} $\bar\beta$. 

Because the possibility of averaging over multiple points in the wings
of the line is practically connected with a dense sampling of
the spectral line, we can again assume the approximate relation 
$2I_\lambda\approx I_{\lambda+\delta\lambda}+I_{\lambda-\delta\lambda}$, and
adopt the simplified form Eq.~(\ref{eq:sgmB_simpl}).
%
%
By approximating $\beta_i\approx\bar\beta$, and recalling that
$2R\approx\lambda_0/\delta\lambda$,
we find
\begin{eqnarray} \label{eq:meanest}
\sigma^2(\bar\beta)
&\approx& 
	\frac{1}{\lambda_0^2}\,\frac{1}{n^2\delta\lambda^2}
	\left(\frac{1}{4R^2}+\frac{1}{2}\,\bar\beta^2\lambda_0^2\right)
	\sum_{i=1}^n
	\frac{I_i}{J^2_i} 
\approx 
	\frac{1}{\lambda_0^2}\,\frac{1}{w^2}
	\left(\frac{1}{4R^2}+\frac{1}{2}\,\bar\beta^2\lambda_0^2\right)
	\frac{1}{J_w^2}
	\sum_{i=1}^n I_i \nonumber \\
&\approx&
	\frac{1}{\lambda_0^2 w^2}
	\frac{\cali_w}{J_w^2}
	\left(\frac{1}{4R^2}+\frac{1}{2}\,\bar\beta^2\lambda_0^2\right)\;,
\end{eqnarray}
where $\cali_w$ is the total photon count over the spectral interval
$w=n\delta\lambda$. 
This expression must be compared with Eq.~(\ref{eq:error-beta}).

We tested the algebraic error estimators (\ref{eq:avgest2}) and (\ref{eq:meanest}) against 
the statistics of magnetic inference via Eq.~(\ref{eq:Vder}) over repeated random realizations of the Poissonian noise. Some results are illustrated in Fig.~\ref{fig:deriv_test}.
The top-left panel of the figure shows a representative line profile
densely sampled ($R\approx 100000$), along with the selected bandpass for 
the application of Eq.~(\ref{eq:Vder}) and the error estimators 
Eqs.~(\ref{eq:avgest2}) and (\ref{eq:meanest}). For
comparison, the top-right panel shows the performance of the error 
estimator Eq.~(\ref{eq:error-beta}) of the integral-moment method, 
similarly to Fig.~\ref{fig:sgmBmoment}.

The two bottom panels demonstrate the magnetic inference errors based on the derivative method, along with the performance of the estimators (\ref{eq:avgest2}) and (\ref{eq:meanest}). In particular, these plots show the standard deviation of the mean magnetic inference (\ref{eq:avgest1}) over 1000 random realizations of the Poissonian noise (star 
symbols), the mean error Eq.~(\ref{eq:avgest2}) with each contributing $\sigma^2(\beta_i)$ being evaluated as the 
pointwise error Eq.~(\ref{eq:sigma2beta}) (blue dots), and finally the
mean-error estimator Eq.~(\ref{eq:meanest}) (orange dots). 
The left panel shows the case of single-point inference (using the 
profile point identified by the blue square in the top-left panel), 
while the right panel shows the case where the errors are averaged 
over all 7 points within the selected bandpass in the blue wing of 
the profile.

When the inference is based on only one wavelength point (bottom-left panel), obviously both 
the mean error Eq.~(\ref{eq:avgest2}) and its estimator Eq.~(\ref{eq:meanest}) coincide with the pointwise error
Eq.~(\ref{eq:sigma2beta}), since $\sigma(\bar\beta)\equiv\sigma(\beta)$. 
In the case of Fig.~\ref{fig:deriv_test}, the expectation value of both
errors calculated over 100 repetitions of the test is 82.9\,G, while
the true standard deviation of the mean inference error (star symbols) gathers around
83.2\,G.

By increasing the number of sampling points used for the magnetic inference to 7 (bottom-right panel), the true standard deviation of the mean gathers around 34.5\,G, which is roughly a factor $\sqrt7$ smaller than for the 1-point derivation.\footnote{Because the ratio $I_\lambda/J^2_\lambda$ entering the 1-point error (see, e.g., Eq.~(\ref{eq:sgmB_simpl})) is not constant across the points used in the average, the expected $1/\sqrt{n}$ scaling is not realized exactly.} The average error is again very well approximated by both Eqs.~(\ref{eq:avgest2}) and (\ref{eq:meanest}), which give 34.5\,G and 34.4\,G, respectively.

Use of both blue and red wings in the spectral-derivative
method is expected to reduce the error by an additional factor $\sqrt2$, e.g., from 34.5\,G down to 24.4\,G. It is worth noting how the integral-moment method (top-right panel) can be considered to perform \emph{comparably} to the 
spectral-derivative approach under the same spectral line modeling assumptions, as it delivers a mean error (28.3\,G) intermediate between the one- and two-wing estimates using Eq.~(\ref{eq:avgest2}).

As a last remark, the use of contiguous spectral bins to improve the error statistics on the magnetic inference error via Eq.~(\ref{eq:avgest2}) raises the question of the effect of signal correlations between neighboring wavelengths, and whether this allows us to treat the various spectral bins as independent, as implied by Eq.~(\ref{eq:avgest2}). We numerically tested the effect of the neglected covariances, and found it to be completely negligible for practical purposes, e.g., leading to a correction of only about 0.05\% to the error predicted by the estimator (\ref{eq:avgest2}) in the examples of Fig.~\ref{fig:deriv_test}.

\section{Conclusions}

We derived algebraic expressions for the errors on the magnetic inference 
via the weak-field approximation (WFA), which can be used to  estimate the magnetic accuracy expected with different instrument designs of spectro-polarimeters, characterized by their throughput (SNR) and spectral resolution. These estimators can therefore be used to inform the 
science-to-instrument requirement flow-down during the design process and
mission planning, enabling a rapid yet reliable assessment of the science specifications 
with regard to magnetic inference that must enter the science-traceability 
matrix of a given instrument, and providing a direct way to verify those 
requirements against predicted instrument performance. 

It must be pointed out (see discussion after Eq.~(\ref{eq:beta}), and references therein) that the WFA assumes that all intensity derivatives are applied to the non-magnetic (hence, non-observable) line profile. All our analytic derivations and numerical verifications have assumed the same. On the other hand, in the application of the
various error estimators to real data, we must substitute such non-observable
$I(\lambda)$ with the observed intensity profile in all the
algebraic expressions of the error estimators, including the various
moments of the intensity, as those errors are determined by the actual photon noise of the observations. Under the characteristic condition of the WFA, $\chi=\Delta\lambda_B/\Delta_T\ll 1$, where $\Delta\lambda_B=\kappa \lambda_0^2 B$ is the Larmor linewidth (see, e.g., \citealt{LL04}; cf.\ also Eqs.~(\ref{eq:ffun})), such a substitution only introduces a relative correction of order $\chi^2$ to the error estimators, and therefore does
not invalidate the applicability of such estimators for the stated purpose of quantifying the impact of magnetic inference errors on the design of polarimetric instruments.

These error estimators were numerically verified against a full statistical analysis of random Monte Carlo realizations of Stokes profiles, under the assumption of Poissonian noise.
For the specific case of the inference of the longitudinal magnetic field strength via the well-known magnetograph formula involving the
Stokes pair $(I,V)$, the error estimators (\ref{eq:sigmabeta}) and (\ref{eq:sigma2beta}) were verified in the numerical examples of Figs.~\ref{fig:sgmBmoment} and \ref{fig:deriv_test}, respectively, using a relatively modest field strength of 100\,G.
%
%
%
Similarly, the inference error estimators for the transverse field projection, for both cases of the integral-moment and spectral-derivative approaches (respectively, Eqs.~(\ref{eq:sigmagam}) and (\ref{eq:sigma2gamma})), were verified numerically in Figs.~\ref{fig:sgmBmomentQ} and \ref{fig:sgmTB}, assuming a stronger field of $1000$\,G. In this case, the numerical tests also helped clarify the well-known problem of the inference bias on the transverse field \cite[e.g.,][]{AR11,Ce18} implied by the quadratic dependence of the linear polarization signals on the field strength through the Zeeman effect. The approach followed in this work was to fully conduct the error analysis on the \emph{unbiased} quantity $\gamma^2$, and convert it to the actual observable $\gamma$ only at the very end via Eq.~(\ref{eq:gam2-gam}). This has allowed us to quantify, both analytically (see App.~\ref{app:bias}) and numerically the effect of the (negative) bias of such estimators, and to identify the regime of SNR values and field strengths that make that bias practically negligible. Nonetheless, it is important to recall that the WFA formula (\ref{eq:Qder}) is generally subject to stricter conditions of applicability than the corresponding Eq.~(\ref{eq:Vder}) for the longitudinal field \citep{LL04}, and therefore, in all practical applications of our error analysis to actual magnetic inference from Stokes data, the true inference errors on transverse fields will include systematic contributions from the breakdown of the WFA, which evidently are not captured by the error estimators derived in this work.

When the wavelength-dependent error estimators Eqs.~(\ref{eq:sigma2beta}) and (\ref{eq:sigma2gamma}) are used to determine the error on the field inclination with respect to the LOS via Eq.~(\ref{eq:sigma2Theta}), Fig.~\ref{fig:sgmTB} clearly shows the strong impact that the larger uncertainty on the determination of $\gamma$ via the WFA for the linear polarization has on the error on the field geometry, practically limiting the domain of applicability to field strengths ${B}\gtrsim 1\,\rm kG$, as well as requiring high-sensitivity measurements with $\rm SNR\gtrsim 10^3$. 
In particular, the right panel of Fig.~\ref{fig:sgmTB} provides a helpful visual of the typical ratio ($\sim 15\times$, for the field geometry considered in that example) between the inference errors $\sigma(B_{\rm POS})$ and $\sigma(B_{\rm LOS})$ that can be expected with full-Stokes Zeeman-effect diagnostics of vector magnetic fields.

The two methods are shown to perform comparably in terms of both the inferred strength and the estimated inference error, when \emph{either} the longitudinal ($\beta$) or the transverse ($\gamma$) component of the field is sought after. In contrast, when both field components are needed simultaneously (e.g., for the inference of the field inclination $\Theta_B$), the derivative method requires distinct, generally non-overlapping spectral windows to achieve that performance (cf.\ Figs.~\ref{fig:sgmTB} and \ref{fig:deriv_test}), whereas the moment method infers both quantities simultaneously from the same, whole-line spectrum, leading to an important operational advantage. Additionally, Fig.~\ref{fig:test_resol} (right panel) shows how the inference errors for the integral-moment approach are nearly independent of the spectral resolution $R$ of the observation, for a fixed total photon count, as long as the line width is sufficiently resolved by the instrument PSF. In that case, the inference error is largely affected by the presence of any background and the associated noise. 
Instead, in the case of the spectral-derivative approach, for a fixed total photon count, Eqs.~(\ref{eq:sgmId}) and (\ref{eq:sgmId2}) show that the noise on the estimated derivatives 
$J_\lambda$ and $K_\lambda$ grows with $R$ as the spectral sampling $\delta\lambda$ is refined. 
This is simply a manifestation of the ill-posedness of the numerical differentiation of noisy data, which can only be mitigated by smoothing the profiles over a fixed spectral scale larger than the sampling width (at the expense of spectral resolution). We note that this is effectively what happens with the integral-moment approach through the integration by parts of Eqs.~(\ref{eq:wfa-moment_V}) and (\ref{eq:wfa-moment_Q}).


Finally, while for all numerical experiments we have adopted specific polarization modulation schemes (e.g., an ideal dedicated $(I,V)$ or $(I,Q)$ dual-beam polarimeter for the examples of Figs.~\ref{fig:sgmBmoment} and \ref{fig:sgmBmomentQ}), it is important to remark that, in the case of weakly polarized targets, the general relations (\ref{eq:v2mom}), (\ref{eq:q2mom}), and (\ref{eq:Stokes_noise}), along with all the algebraic error estimators that rely on them, remain applicable to any type of modulation scheme characterized by a given set of $r_{Q,U,V}$ values. For arbitrarily polarized targets, the same estimators also remain applicable for particular types of polarimeters, which, however, only need to satisfy rather broad constraints commonly addressed (and approximately met) by existing designs of polarimetric instruments (see App.~\ref{app:covariances}).

In summary, the numerical results shown in
this work demonstrate both the applicability and the robustness of the algebraic error estimators we derived, and also how the integral-moment and spectral-derivative approaches compare quantitatively as to their inference errors.

\acknowledgements

This material is based upon work supported by the NSF
National Center for Atmospheric Research, which is a major facility sponsored by the National
Science Foundation under Cooperative Agreement No. 1852977. The author acknowledges helpful discussions on the subject of this work with R.\ Centeno (NSF NCAR, HAO) and J.~C.\ del Toro Iniesta (IAA, Spain). The comments of the editor and referee were also extremely useful to point out argument weaknesses and unspotted errors, and to significantly improve the presentation.

During the development of this work, Anthropic Claude (Opus 5.5) was used to help verify the analytic derivations done by the author, either directly or with the help of algebraic symbolic programs (wxMaxima), to quantify critical covariance terms, and to help debug or augment the capabilities of the IDL codes the author has programmed to implement the formalism and produce the figures; the entire narrative, all numerical results, and the conclusions of this study were the exclusive product of the author.

\begin{appendix}

\section{Stokes covariances for a general modulation scheme}
\label{app:covariances}

In this appendix, we derive the covariances among Stokes parameters adopted in the article, assuming a general polarization modulation scheme with $n$ states. We follow the notation of Sect.~2 of \citeauthor{Ca26} (\citeyear{Ca26}; hereafter CHdW), where a generalized definition of modulation efficiency was recently proposed. 

In a given spectral bin, the Stokes vector of
the target is $\bm{S}=(S_1,S_2,S_3,S_4)^t\equiv(I,Q,U,V)^t$, and
the $n$-vector of modulated signals is
$\bm{I}=\mathbf{O}\bm{S}=\mathbf{T}\mathbf{\tilde{O}}\bm{S}$,
where the diagonal matrix $\mathbf{T}=(\delta_{ij} t_j)$ contains the transmissivities of the $n$ modulation states, thus, $O_{i1}=t_i$ and $\tilde{O}_{i1}=1$.
The Stokes vector of the target is inferred via $\bm{S}=\mathbf{D}\bm{I}$,
where
$\mathbf{D}=\bm{\lambda}\mathbf{O}^t\mathbf{T}^{-1}=\bm{\lambda}\mathbf{\tilde{O}}^t$ is the optimal demodulation matrix, with
$\bm{\lambda}=(\mathbf{O}^t\mathbf{T}^{-1}\mathbf{O})^{-1}=(\mathbf{\tilde{O}}^t\mathbf{T}\mathbf{\tilde{O}})^{-1}=\mathbf{D}\mathbf{T}\mathbf{D}^t$  
(see CHdW, Eqs.~(7) and (8)). The modulated signals $I_i$ are assumed to be uncorrelated and photon-noise limited, so that $\sigma^2(I_i)\approx k I_i$ (CHdW, footnote~3). In particular, if all signals are expressed in photon counts, we can simply set $k=1$. We assume this is the case for the following development.

The elements of the covariance matrix between the demodulated Stokes signals are
\begin{eqnarray} \label{eq:Scov}
\mathrm{Cov}(S_i,S_j)
&=&\sum\nolimits_l D_{il}D_{jl}\,\sigma^2(I_l)
\approx \sum\nolimits_l D_{il} I_l D^t_{lj} \nonumber \\
&\approx&\sum\nolimits_{lk}
D_{il} (\delta_{lk} I_k) D^t_{kj}
\equiv(\mathbf{D}\mathbf{I}\mathbf{D}^t)_{ij}
=(\bm{\lambda}\mathbf{\tilde{O}}^t
\mathbf{I}
\mathbf{\tilde{O}}\bm{\lambda})_{ij}\;,
\end{eqnarray}
where in the last row we also introduced the diagonal matrix $\mathbf{I}=(\delta_{ij}I_j)$ of the modulated signals.
We note that Eq.~(4) of CHdW corresponds to the diagonal elements of Eq.~(\ref{eq:Scov}) under the approximations of that previous work.

Following CHdW, the diagonal elements of $\bm{\lambda}$ are simply proportional to the inverse squares of the modulation efficiencies, $n\lambda_{ii}\propto\epsilon_i^{-2}$. In particular,
\begin{equation} \label{eq:rSdef}
\frac{\lambda_{ii}}{\lambda_{11}}=\frac{\epsilon_1^2}{\epsilon_i^2}\equiv r_{S_i}^2\;,
\end{equation}
where in the last equivalence we recalled the notation introduced in this paper (see, e.g., Eq.~(\ref{eq:q2mom})). It is important to remark that various modulation schemes commonly adopted in applied spectro-polarimetry are ideally characterized by a diagonal $\bm{\lambda}$ matrix \cite[see, e.g.,][Eqs.~(18) and (20)]{dTIC00}. For example, this is true for the typical dual-beam polarimeter design based on a rotating stack of retarders sampling at equally spaced angular positions over the 180$^\circ$ period.
In particular, if a modulation scheme is characterized by a diagonal $\bm{\lambda}$ matrix, from the definition of $\mathbf{D}$ we see that $D_{1j}=\lambda_{11}$ for all modulation states $j$. In fact, for such a condition to be satisfied, it is sufficient for $\bm{\lambda}$ to only be \emph{block-diagonal} with respect to its intensity and polarization parts, i.e., $\lambda_{1j}=\lambda_{j1}=0$, even if the polarization block of $\bm{\lambda}$ may not be diagonal.

In order to apply Eq.~(\ref{eq:Scov}) to the development presented in the main text, we consider below different levels of approximations relevant to this work.

\subsection{Weakly polarized target}

%

When the light from the target is weakly polarized (WP), we can assume $I_i\approx t_i S_1$ for all modulation states, so we can replace $\mathbf{I}=S_1\mathbf{T}$ in Eq.~(\ref{eq:Scov}).
In that case, the elements of the 
covariance matrix of Eq.~(\ref{eq:Scov}) become
\begin{equation} \label{eq:ScovWP} 
\mathrm{Cov}(S_i,S_j)_\mathrm{WP}
\approx S_1\,(\bm{\lambda}\mathbf{\tilde{O}}^t\mathbf{T}\mathbf{\tilde{O}}\bm{\lambda})_{ij}
\approx\lambda_{ij}\,S_1\;,
\end{equation}
where we recalled that $\mathbf{\tilde{O}}^t\mathbf{T}\mathbf{\tilde{O}}=\bm{\lambda}^{-1}$. 
The above relation applies independently at each spectral bin. 
We thus see that, under the WP assumption, all cross-Stokes covariances vanish \emph{at all wavelengths} in the case of a modulation scheme characterized by a diagonal $\bm{\lambda}$ matrix.

In order to use Eq.~(\ref{eq:ScovWP}) to calculate the covariance between the integral moments of orders $p$ and $q$ of any two Stokes parameters, $S_i$ and $S_j$, according to the definition (\ref{eq:integral}), it is sufficient to observe that 
the statistical noise of each modulated signal is independent between spectral bins, and that the moment weights simply multiply in the 
covariance integral. Therefore, we get at once
\begin{equation} \label{eq:momcov}
  \mathrm{Cov}\bigl( (\cals_i)_p,(\cals_j)_q\bigr)_\mathrm{WP}
  \approx\lambda_{ij}\,\cali_{p+q}\;,
\end{equation}
where, e.g., $(\cals_1)_p\equiv\cali_p$, $(\cals_2))_q\equiv\calq_q$,
etc. 
In particular, for the covariances that we needed to consider to evaluate the variance estimators (\ref{eq:sigmabeta}), (\ref{eq:sigmagam}), and (\ref{eq:TB}), we find
\begin{equation} \label{eq:WPcov}
\mathrm{Cov}\bigl(\calv_1,\cali_0\bigr)_\mathrm{WP}\approx\lambda_{41}\,\cali_1\;,\quad
\mathrm{Cov}\bigl(\calq_2,\cali_0\bigr)_\mathrm{WP}\approx\lambda_{21}\,\cali_2\;,\quad
\mathrm{Cov}\bigl(\calv_1,\calq_2\bigr)_\mathrm{WP}\approx\lambda_{42}\,\cali_3\;.
\end{equation}
The two covariances involving Stokes $V$ are found to vanish in the case of a symmetric line, because of the vanishing of the odd moments of the intensity. The covariance involving Stokes $Q$ instead does not vanish under the same argument, being driven by the total width of the line; in addition, single-beam polarimeters implementing a rotating stack of waveplates naturally have $\lambda_{12}=\lambda_{21}\ne 0$.
%
This helps clarify the conditions under which the covariances adopted in this work can be neglected or must be retained.

We must note that in Sect.~\ref{sec:moments} we implicitly assumed the inferred Stokes intensity to be Poisson distributed, in order to write Eq.~(\ref{eq:noise1}); similarly, in Sect.~\ref{sec:deriv}, we consistently assumed $\sigma^2(I_\lambda)\approx I_\lambda$ for all Stokes intensity signals. However, the Stokes intensity inferred via demodulation of $\bm{I}=\mathbf{O}\bm{S}$,
\begin{equation} \label{eq:StkI}
S_1=\sum\nolimits_j D_{1j} I_j\;,
\end{equation}
in general is \emph{not} Poisson distributed, as
\begin{equation} \label{eq:sgmStkI}
\sigma^2(S_1)\approx \sum\nolimits_j D_{1j}^2\sigma^2(I_j)
\approx \sum\nolimits_j D_{1j}^2 I_j
\not\approx S_1\;.
\end{equation}
This is evidently the case because Eq.~(\ref{eq:StkI}) corresponds to a weighted average of the modulated signals $I_i$.
%
On the other hand, in Sects.~\ref{sec:moments} and \ref{sec:deriv} we exclusively relied on a condition of equality between the variance and the mean of the inferred Stokes intensity, so a strict characterization of $S_1$ as a Poisson distributed quantity is actually not necessary (albeit sufficient) for our needs.

It is a remarkable result implied by Eq.~(\ref{eq:ScovWP}) that a simple normalization of the demodulation matrix by $\lambda_{11}$ is sufficient to realize the desired correspondence of variance and mean, \emph{under the assumption of weakly polarized targets.} 
As weakly polarized targets are commonly encountered in astronomical spectro-polarimetry, such a normalization appears to be a sensible approach.
To appreciate the physical meaning of such a normalization, we note that $I_j\approx t_j S_1$ implies
\begin{equation} \label{eq:StkI.alt}
\sum\nolimits_j I_j\approx S_1\sum\nolimits_j t_j=n \bar{t}\,S_1\;,
\end{equation}
where $\bar{t}$ is the average transmissivity of the modulation scheme, and comparing Eqs.~(\ref{eq:StkI}) and (\ref{eq:StkI.alt}) in the case of a block-diagonal $\bm{\lambda}$ matrix shows that $D_{1j}=\lambda_{11}=(n\bar{t})^{-1}$. Thus, the normalization of $\bm{\lambda}$ by $\lambda_{11}$ corresponds approximately to rescaling the Stokes intensity by $n\bar{t}$, such a correspondence becoming exact in the case of a block-diagonal $\bm{\lambda}$, in which case the scaled intensity (\ref{eq:StkI.alt}) actually becomes Poisson distributed. 

We note that we can attain the suggested rescaling (\ref{eq:StkI.alt}) by orderly modifying the definitions of the modulation and demodulation matrices according to $\mathbf{O}'=\mathbf{O}/\bar{t}$ and $\mathbf{D}''=n\mathbf{D}'$, so that $\bm{\lambda}''=n\bar{t}\,\bm{\lambda}$. The rescaling of $\mathbf{D}$ directly avoids the averaging of the modulated signals, whereas the rescaling of $\mathbf{O}$ is always possible because the determination of the modulation matrix via polarimetric calibration only fixes $\mathbf{O}$ up to a constant factor. In fact, in the field of astronomical spectro-polarimetry, it has been common practice to normalize $\mathbf{O}$ by setting the largest value of the first column to 1 (see, e.g., CHdW, Eq.~(17)). On the other hand, such a choice leaves behind a spurious gain factor in Eq.~(\ref{eq:StkI.alt}) that prevents the rescaled Stokes intensity from behaving like a Poisson-distributed quantity when $\lambda_{11}=(n\bar{t})^{-1}$. Here we have shown instead that the normalization of $\mathbf{O}$ by the average transmissivity is a more sensible choice to approach the desired Poissonian character of the rescaled $S_1$. 

In the following, we assume that the normalization of $\bm{\lambda}$ by $\lambda_{11}$ has been imposed from the start, so $\lambda_{11}=1$ and $\lambda_{ii}=r_{S_i}^2$.
With this normalization, we see that Eqs.~(\ref{eq:noise1}), (\ref{eq:v2mom}), and (\ref{eq:q2mom}) in the paper follow directly from Eq.~(\ref{eq:momcov}).

%
%



\subsection{Covariances to the lowest order of polarization}

The cross-Stokes covariances of Eq.~(\ref{eq:WPcov}), which may be vanishing under the WP approximation considered above, were found nonetheless to be relevant to the estimators of Eqs.~(\ref{eq:sigmabeta}), (\ref{eq:sigmagam}), and (\ref{eq:TB}). In fact, while such covariances may only be non-vanishing because of contributions that are of first order in the polarization, they enter the various variance estimators through rational terms of the form (\ref{eq:EPrelgen}), which contribute to such estimators at the same level as the intensity terms. 

Because Eq.~(\ref{eq:WPcov}) already comprises the dominant contributions from non-diagonal elements of $\bm{\lambda}$, here we restrict our analysis to a modulation scheme with a diagonal $\bm{\lambda}$ matrix. In this case, we can rewrite Eq.~(\ref{eq:Scov}) more simply as
\begin{equation} \label{eq:Scov-diag}
\mathrm{Cov}(S_i,S_j)
\approx \lambda_{ii}\lambda_{jj}
(\mathbf{\tilde{O}}^t
\mathbf{I}
\mathbf{\tilde{O}})_{ij}
\approx r_{S_i}^2 r_{S_j}^2(\mathbf{\tilde{O}}^t\mathbf{I}\mathbf{\tilde{O}})_{ij}\;.
\end{equation}
We consider first the case of the intensity-polarization covariances, and recall Eq.~(\ref{eq:Scov}) to rewrite ($r_I=1$)
\begin{equation} \label{eq:S2Icov}
\mathrm{Cov}(S_1,S_j)
\approx r^2_{S_j}
\sum\nolimits_l \tilde{O}^t_{1l} I_l \tilde{O}_{lj}
\approx r^2_{S_j}
\sum\nolimits_l \tilde{O}^t_{jl} I_l
=S_j
\end{equation}
(since $\tilde{O}_{1l}^t=1$).
Therefore, to the lowest order of polarization,  Eq.~(\ref{eq:momcov}) for $\cals_i=\cali$ acquires a correction given by
\begin{equation} \label{eq:polcov}
\mathrm{Cov}(\cali_p,(\cals_j)_q)
\approx(\cals_j)_{p+q}\;.
\end{equation}
These covariances were taken into account in the expressions of the variance estimators (\ref{eq:sigmabeta}) and (\ref{eq:sigmagam}). When we also set $\cals_j=\cali$, Eq.~(\ref{eq:polcov}) confirms once again the validity of the approximation (\ref{eq:noise1}) used in the derivation of the error estimators presented in this work. 

In general, it is important to remark that an arbitrary polarimeter with a non-diagonal $\bm{\lambda}$ will produce polarization dependent corrections to the Stokes variances (i.e., for $i=j$ in Eq.~(\ref{eq:Scov})), which affect Eq.~(\ref{eq:noise1}) as well as the derived relations (\ref{eq:v2mom}) and (\ref{eq:q2mom}). For mild deviations from diagonality (e.g., as they may be caused by a $\lesssim$\,10\% variation of the system's throughput across the modulation states), these corrections are at most only a few percent even in the presence of strong polarization, and they can safely be ignored. More importantly, under the WFA assumption, since all polarization signals exclusively arise from the Zeeman effect, their zeroth-order moments vanish identically when the spectral range of the line reaches out to the neighboring continuum, and in particular the fundamental relation $\sigma^2(\cali_0)\approx\cali_0$ adopted throughout Sect.~\ref{sec:moments} remains generally applicable even for an arbitrary polarimeter. On the other hand, variances that are estimated at different wavelength positions along the spectral line (e.g., Eq.~(\ref{eq:Stokes_noise})) are generally susceptible to corrections induced by the non-diagonality of $\bm{\lambda}$ and would need to be assessed on a case-by-case basis.

Equation~(\ref{eq:Scov-diag}) also gives rise to non-vanishing covariance terms such as
$\mathrm{Cov}(\calv_1,\calq_2)$, which must be taken into account for the inference of the magnetic field geometry (see Eqs.~(\ref{eq:TB}) and (\ref{eq:sigma2Theta})). 
In order to estimate the importance of these terms, we assume the ideal case of a single-beam polarimeter with modulation matrix
\begin{equation} \label{eq:tetra}
\mathbf{\tilde{O}}=
\begin{pmatrix}
1 &r_Q^{-1} &r_U^{-1} &r_V^{-1} \\
1 &r_Q^{-1} &-r_U^{-1} &-r_V^{-1} \\
1 &-r_Q^{-1} &r_U^{-1} &-r_V^{-1} \\
1 &-r_Q^{-1} &-r_U^{-1} &r_V^{-1}
\end{pmatrix}\;,
\end{equation}
and uniform transmissitivity matrix $\mathbf{T}=t\,\mathbf{1}$. We note that such a modulator possesses a diagonal $\bm{\lambda}$ matrix, although it is not necessarily maximally efficient, unless $r_Q^{-2}+r_U^{-2}+r_V^{-2}=1$.
In this case, we find
\begin{equation} \label{eq:QVcov}
\mathrm{Cov}(V,Q)\approx \frac{r_Q\,r_V}{r_U}\,U\;,\quad
\mathrm{Cov}(\calv_1,\calq_2)\approx \frac{r_Q\,r_V}{r_U}\,\calu_3\;,
\end{equation}
and for the typical case of a symmetric Stokes $U$, the moment covariance vanishes. Instead, the wavelength dependent covariance on the left may contribute significantly to the estimator of Eq.~(\ref{eq:sigma2Theta}).
Both covariances (\ref{eq:QVcov}) are instead found to vanish identically for a perfectly balanced dual-beam polarimeter based on the same modulation scheme (\ref{eq:tetra}). In the case of imperfect beam balance, these covariances are non-vanishing and scale with the same degree of beam imbalance $\rho$ introduced by \cite{Ca12} for modeling seeing-induced cross-talk in dual-beam polarimeters. However, unlike polarization cross-talk, in this case the effect of beam imbalance on the covariances (\ref{eq:QVcov}) cannot be removed by rebalancing the beams during data reduction, as it is driven by the noise of the acquired signals rather than their amplitude.


It is worth noting how the common design of a dual-beam polarimeter employing a rotating birefringent stack, and sampling at equally spaced positions over the 180$^\circ$ period, is resilient to the appearance of various covariance terms. When the two beams are perfectly balanced, the $\bm{\lambda}$ matrix of such a system is fully diagonal when the birefringent stack acts as a linear retarder. A circular component in the retardance couples the $Q$ and $U$ elements of $\bm{\lambda}$, introducing a corresponding covariance of the form (\ref{eq:ScovWP}). However, the $\bm{\lambda}$ matrix remains block-diagonal, and in particular the linear-to-circular off-diagonal elements remain null. In either case, all first-order covariances between the polarization parameters vanish. The other covariances that survive are the intensity–polarization terms of Eqs.~(\ref{eq:S2Icov}) and (\ref{eq:polcov}), which are already included in the estimators derived in the main text. Any imbalance between the two beams reintroduces the covariances between polarization parameters, of the order of the degree of beam imbalance $\rho$.

\section{Quantification of the error bias on $\sigma(\gamma)$}
\label{app:bias}

Let $v$ be a random Gaussian variable, and $\bar{v}=E(v)$ its expectation value. We note that we can write
\begin{equation} \label{eq:v}
v=\bar{v}(1+\epsilon)\;,\quad
\epsilon\equiv\frac{v-\bar{v}}{\bar{v}}\;,
\end{equation}
where $\epsilon$ is also a random Gaussian variable, such that
\begin{equation} \label{eq:eps}
E(\epsilon)=0\;,\quad
\sigma^2_\epsilon\equiv E(\epsilon^2)=\frac{\sigma^2(v)}{\bar{v}^2}\;.
\end{equation}

Under the rather loose requirement that $|\epsilon|<1$, from Eq.~(\ref{eq:v}) we have
\begin{equation}
\sqrt{v}=\sqrt{\bar{v}}\,(1+\epsilon)^{1/2}
\approx\sqrt{\bar{v}}\left(
1+\frac{1}{2}\epsilon-\frac{1}{8}\epsilon^2+\frac{1}{16}\epsilon^3-\frac{5}{128}\epsilon^4+\cdots\right)
\equiv\sqrt{\bar{v}}\,(1+\delta)\;,
\end{equation}
and thus
\begin{equation}
\sigma^2(\sqrt{v})=\bar{v}\,\sigma^2(\delta)
=\bar{v}\left[E(\delta^2)-E(\delta)^2\right]\;.
\end{equation}

The expectation value of $\delta$ is easily calculated recalling the expression for the moments of a normal random variable \cite[see, e.g.,][Eq.~(5-47)]{Pa65}
\begin{equation}
E(\epsilon^{2k-1})=0\;,\quad
E(\epsilon^{2k})=(2k-1)!!\,\sigma_\epsilon^{2k}\;,\qquad k\in\mathbb{N}\;,
\end{equation}
so that
\begin{eqnarray*}
E(\delta)
&=&-\frac{1}{8}\sigma_\epsilon^2
-\frac{15}{128}\sigma_\epsilon^4+\cdots\;, \\
E(\delta^2)
&=&E\biggl(
\frac{1}{4}\epsilon^2
-\frac{1}{8}\epsilon^3
+\frac{5}{64}\epsilon^4+\cdots\biggr)
=\frac{1}{4}\sigma_\epsilon^2
+\frac{15}{64}\sigma_\epsilon^4+\cdots\;,
\end{eqnarray*}
from which
\begin{displaymath}
E(\delta^2)-E(\delta)^2=\frac{1}{4}\sigma_\epsilon^2+\frac{7}{32}\sigma_\epsilon^4+\cdots
=\frac{1}{4}\sigma_\epsilon^2\left(1+\frac{7}{8}\sigma_\epsilon^2+\cdots\right)\;,
\end{displaymath}
and finally
\begin{equation} \label{eq:last}
\sigma^2(\sqrt{v})=\frac{1}{4}\frac{\sigma^2(v)}{\bar{v}}\left(1+\frac{7}{8}\frac{\sigma^2(v)}{\bar{v}^2}+\cdots\right)\;.
\end{equation}

In our case, $v=\gamma^2$, and the error expression (\ref{eq:gam2-gam}) thus corresponds to the lowest order of approximation of Eq.~(\ref{eq:last}), which is applicable when $\sigma(\gamma^2)/\gamma^2\ll 1$. When such a condition is not strictly met, the approximation of Eq.~(\ref{eq:gam2-gam}) provides a biased estimate of the error on $\gamma$, which \emph{always underestimates} the true random error. This is clearly illustrated by Fig.~\ref{fig:sgmBmomentQ}, where in the case of the left panel the second-order correction to the error is about 2.4\%, whereas in the case of the right panel it amounts to about 0.6\%. Using successive orders in Eq.~(\ref{eq:last}) allows us to make the error estimator more accurate.

\end{appendix}

\bibliographystyle{apalike}
\bibliography{mn}

\end{document}